\documentclass[10pt,a4paper,twocolumn,english,prb,aps,showpacs,floatfix,groupedaddress,superscriptaddress]{revtex4-1}
\usepackage{graphicx,epsfig}
\usepackage{times}
\usepackage{graphics,dcolumn,bm,float}
\usepackage{amssymb,amsmath,rotate,color}
\usepackage[breaklinks,colorlinks=true,urlcolor=blue,citecolor=blue,linkcolor=blue]{hyperref}
\usepackage{breakurl}

\begin{document}
\unitlength 1 cm
\newcommand{\be}{\begin{equation}}
\newcommand{\ee}{\end{equation}}
\newcommand{\bearr}{\begin{eqnarray}}
\newcommand{\eearr}{\end{eqnarray}}
\newcommand{\nn}{\nonumber}
\newcommand{\dagg}{{\dagger}}
\newcommand{\vpdag}{{\vphantom{\dagger}}}
\newcommand{\vecr}{\vec{r}}
\newcommand{\bs}{\boldsymbol}
\newcommand{\up}{\uparrow}
\newcommand{\down}{\downarrow}
\newcommand{\fns}{\footnotesize}
\newcommand{\ns}{\normalsize}
\newcommand{\cdag}{c^{\dagger}}

\definecolor{red}{rgb}{1.0,0.0,0.0}
\definecolor{green}{rgb}{0.0,1.0,0.0}
\definecolor{blue}{rgb}{0.0,0.0,1.0}

\title{Lattice symmetry and emergence of antiferromagnetic quantum Hall states}

\author{Morad Ebrahimkhas}
\email{ebrahimkhas@iau-mahabad.ac.ir}
\affiliation{Department of Physics, Mahabad Branch, Islamic Azad University, Mahabad , Iran}
\author{Mohsen Hafez-Torbati}
\email{mohsen.hafez@tu-dortmund.de}
\affiliation{Institut f\"ur Theoretische Physik, Goethe-Universit\"at, 60438 Frankfurt/Main, Germany}
\affiliation{Lehrstuhl f\"ur Theoretische Physik I, 
Technische Universit\"at Dortmund,
Otto-Hahn-Stra\ss e 4, 44221 Dortmund, Germany}
\author{Walter Hofstetter}
\email{hofstett@physik.uni-frankfurt.de}
\affiliation{Institut f\"ur Theoretische Physik, Goethe-Universit\"at, 60438 Frankfurt/Main, Germany}

\begin{abstract}
Strong local interaction in systems with non-trivial topological bands can stabilize quantum 
states such as magnetic topological insulators.
We investigate the influence of the lattice symmetry on the possible 
emergence of antiferromagnetic quantum Hall states.
We consider the spinful Harper-Hofstadter model extended by a next-nearest-neighbor (NNN) 
hopping which opens a gap at half-filling and allows for the realization of a quantum Hall insulator. 
The quantum Hall insulator has the Chern number $\mathcal{C}=2$ as both spin components are 
in the same quantum Hall state. We add to the system a 
staggered potential $\Delta$ along the $\hat{x}$-direction 
favoring a normal insulator and the Hubbard interaction $U$ favoring a Mott insulator. 
The Mott insulator is a N\'eel antiferromagnet for small and a 
stripe antiferromagnet for large NNN hopping. 
We investigate the $U$-$\Delta$ phase diagram of the model for both small and large NNN hoppings. 
We show that while for large NNN hopping there exists a $\mathcal{C}=1$ stripe antiferromagnetic 
quantum Hall insulator 
in the phase diagram, there is no equivalent $\mathcal{C}=1$ N\'eel antiferromagnetic quantum 
Hall insulator at the small NNN hopping. 
We discuss that a $\mathcal{C}=1$ antiferromagnetic quantum Hall insulator can emerge only if 
the effect of the spin-flip transformation cannot be compensated by a space group operation. 
Our findings can be used as a guideline in future investigations searching for 
antiferromagnetic quantum Hall states.
\end{abstract}

\maketitle

\section{Introduction}
The role of symmetry in the development of modern condensed matter 
physics especially in the field of topological insulators (TIs) is unequivocally recognized. 
Magnetic TIs characterized by a non-trivial topological invariant and long-range 
magnetic order are promising candidates for application in dissipationless 
quantum transport, low-energy consumption 
spintronics, and topological quantum computation \cite{Tokura2019}.
The recent realization of  MnBi$_2$Te$_4$ as the first antiferromagnetic TI 
has been a key advance in the field of magnetic TIs 
\cite{Zhang2019,Otrokov2019a,Li2019a,Li2019,Otrokov2019}.

The experimental achievements in creation of artificial gauge fields \cite{Cooper2019,Aidelsburger2018}
and in detection of magnetic order \cite{Mazurenko2017,Brown2017} 
suggest ultracold atoms trapped in optical lattices \cite{Hofstetter2018} as a unique 
system for simulating magnetic topological quantum states with a high degree of control and tunability of parameters.
The Harper-Hofstadter model is realized in optical lattices 
using the laser-assisted-tunneling \cite{Aidelsburger2013,Miyake2013}.  
The Haldane model is implemented using the lattice-shaking technique \cite{Jotzu2014}. 
Further developments are measuring the Chern number of the Hofstadter bands \cite{Aidelsburger2014}
and the Berry curvature of the Bloch bands \cite{Flaschner2016}.

Feshbach resonances can be used to tune the interaction between ultracold atoms \citep{Bloch2008}. 
The effect of interaction on topological systems has become an interesting problem 
in recent years \cite{Rachel2018}. 
In the spinless Haldane model the nearest-neighbor interaction induces a transition from 
a Chern insulator to a charge ordered Mott insulator (MI) \cite{Varney2010}. 
In spinful systems the Hubbard 
interaction can drive a normal insulator (NI) into a 
quantum Hall \cite{He2011,Vanhala2016,Tupitsyn2019,Mertz2019} 
or quantum spin 
Hall insulator \cite{Cocks2012,Budich2013,Amaricci2015,Jiang2018}. 
Interaction-driven topological transitions are studied also in three-dimensional systems \cite{Amaricci2016,Irsigler2020}.
In SU(3) systems, topological transitions from a magnetic insulator into a quantum 
Hall insulator (QHI) are reported which have no counterparts in the SU(2) 
case \cite{Hafez-Torbati2020}. 

In the strong coupling limit the Hubbard interaction favors 
long-range magnetic order, unless quantum fluctuations are 
strong enough to stabilize a quantum spin liquid or a valence 
bond crystal state \cite{Balents2010}. This can lead to novel magnetic orders when 
artificial gauge fields or spin-orbit coupling are present in the 
system \cite{Cocks2012,Radic2012,Arun2016,Irsigler2019a}. 
In addition, the competition between the band insulator at weak and the Mott insulator
at strong interaction can stabilize novel intermediate phases such as antiferromagnetic QHI (AFQHI) 
with Chern number $\mathcal{C}=1$ as suggested for 
the Haldane-Hubbard model \cite{He2011,Vanhala2016,Tupitsyn2019}.
In this phase, one of the spin components is in the quantum Hall state and the other in the 
normal state.

In this paper 
we investigate whether the $\mathcal{C}=1$ AFQHI is a phase specific to the Haldane-Hubbard model 
or whether it can occur in other interacting topological systems. 
With this aim we consider the spinful Harper-Hofstadter model in the presence of 
the Hubbard interaction $U$, i.e. the Harper-Hofstadter-Hubbard model, at half-filling with 
the plaquette magnetic flux $1/2$ in units of magnetic flux quantum $h/e$. 
The flux is the same for both spin components. 
The Harper-Hofstadter model at half-filling is gapless and hence we add a next-nearest-neighbor (NNN) 
hopping to the system to open a gap and realize a QHI \cite{Hatsugai1990}. 
The QHI has the Chern number $\mathcal{C}=2$. 
The $\mathcal{C}=1$ AFQHI in the Haldane-Hubbard model 
appears as a result of competition between the staggered potential and the Hubbard 
interaction \cite{He2011,Vanhala2016,Tupitsyn2019}. 
We include in our model also a staggered potential $\Delta$ which favors a NI phase.

\section{model and main results}
The Hamiltonian of the system reads
\be
H=H_t+\Delta \sum_{\vec{r},\sigma}(-1)^x n^\vpdag_{\vec{r},\sigma}
+U\sum_{\vec{r}} n^\vpdag_{\vec{r},\down} n^\vpdag_{\vec{r},\up}
\label{4h.eq}
\ee
with the hopping term 
\begin{eqnarray}
\label{eq:hopping}
\noindent
   H_t&=&-\sum_{\vec{r},\sigma}
   \left( tc^\dagger_{\vec{r}+\hat{x},\sigma}c^\vpdag_{\vec{r},\sigma}
   +te^{2\pi i \varphi x} c^\dagger_{\vec{r}+\hat{y},\sigma}c^\vpdag_{\vec{r},\sigma} \right.
   + t'\times \phantom{++} \nn \\
   &&\hspace{-0.6cm}\left. e^{2\pi i \varphi (x+1/2)} (c^\dagger_{\vec{r}+\hat{x}+\hat{y},\sigma}
   c^\vpdag_{\vec{r},\sigma}+ c^\dagger_{\vec{r}+\hat{y},\sigma}
   c^\vpdag_{\vec{r}+\hat{x},\sigma})+{\rm H.c.} \right) 
\end{eqnarray}
where $t$ and $t'$ are the NN and the NNN hoppings, respectively.
The fermionic operator $c^\dagger_{\vec{r},\sigma}$ ($c^\vpdag_{\vec{r},\sigma}$) creates (annihilates) a 
particle at position $\vec{r}=x\hat{x}+y\hat{y}=(x,y)$ with spin component $\sigma=\up,\down$. The position $\vec{r}$ 
runs over the square lattice and the lattice constant is considered as the unit of length. We define the occupation
number operator $n_{\vecr,\sigma}^\vpdag=c^\dagg_{\vecr,\sigma}c^\vpdag_{\vec{r},\sigma}$.
The parameter $\varphi$ is the magnetic flux entering each square, in units of the magnetic flux quantum. 
We fix $\varphi=1/2$ which is the simplest flux in the Harper-Hofstadter model 
to achieve topological bands. We would like to point out that other fluxes such as $1/4$ can 
stabilize quantum Hall states with higher Chern numbers \cite{Hatsugai1990} and are also interesting 
to investigate.

The effect of the NNN hopping on the Harper-Hofstadter model is studied in a number 
of previous work \cite{Hatsugai1990,Han1994,Thouless1983}. It is included in the 
Hamiltonian Eq. \eqref{4h.eq} to open a gap at half-filling and realize a QHI \cite{Hatsugai1990}.
The ratio of the NNN hopping to the NN hopping in optical lattices can be tuned from weak to strong 
using the lattice shaking technique \cite{Beugeling2012,DiLiberto2011}.

The second term in Eq. \eqref{4h.eq} is a staggered potential along $\hat{x}$-direction, with sublattices 
$A$ and $B$ acquiring, respectively, the onsite energies $+\Delta$ and $-\Delta$. 
Such a staggered potential allows a NI to appear in the phase diagram. It can be easily created 
in optical lattices and is present in the suggested experimental setups \cite{Gerbier2010,Goldman2010}. 
Another possibility would be the checkerboard potential which yields an energy offset 
between the lattice sites with $x+y$ even and the lattice sites with $x+y$ odd.
The last term is the Hubbard interaction.

\begin{figure}[t]
   \begin{center}
   \includegraphics[width=0.4\textwidth,angle=0]{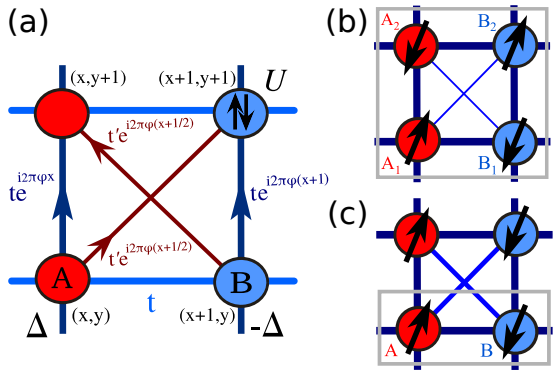}
   \caption{(a) Schematic representation of the Hamiltonian Eq. \eqref{4h.eq}. 
   Schematic representation of the N\'eel (b) and the stripe (c) antiferromagnet with 
   the gray box specifying the unit cell.}
   \label{4H.fig}
   \end{center}
\end{figure}

Our proposed model Eq. \eqref{4h.eq} is the minimal extension of the 
Harper-Hofstadter-Hubbard model which 
allows to examine the existence of a $\mathcal{C}=1$ AFQHI beyond the Haldane-Hubbard model.
One notes that we are considering artificial gauge fields \cite{Cooper2019, Aidelsburger2018}, 
which is why no Zeeman term exists in the Hamiltonian Eq. \eqref{4h.eq}.

The Hamiltonian is schematically depicted in Fig. \ref{4H.fig}(a).
For $U=0$ the Hamiltonian reduces to a two-level problem in momentum space and for finite $t'$ leads to 
a transition between the QHI and the NI 
at $\Delta=2t$ \cite{Zheng2019}. 
If there is no flux and no NNN hopping the Hamiltonian reminisces the ionic Hubbard model 
with a NI for weak and a N\'eel AF for strong $U$. 
There are suggestions for intermediate phases 
\cite{Hafez-Torbati2016,Kancharla2007,Paris2007,Wang2020,Lin2015,Ebrahimkhas2011,Shahbazy2019}.

\begin{figure}[t]
   \begin{center}
   \includegraphics[width=0.4\textwidth,angle=0]{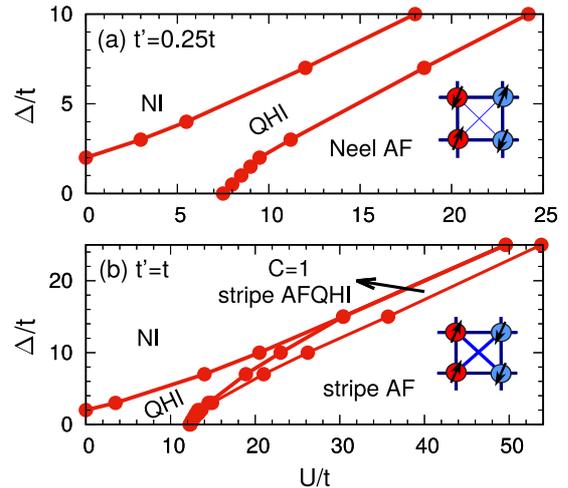}
   \caption{The phase diagram of the Hamiltonian Eq. \eqref{4h.eq} for $\varphi=1/2$ with
   next-nearest-neighbor hopping $t'=0.25t$ (a) and $t'=t$ (b). One can identify normal insulator (NI), 
   $\mathcal{C}=2$ quantum Hall insulator (QHI), N\'eel and stripe antiferromagnet (AF), and a $\mathcal{C}=1$
   stripe antiferromagnetic QHI (AFQHI) in the phase diagram.}
   \label{fig:phase_diagram}
   \end{center}
\end{figure}

We study the phase diagram of the model Eq. \eqref{4h.eq} 
in the $U$-$\Delta$ plane both for small and for large NNN hopping,
in units of nearest-neighbor (NN) hopping $t$. 
The results are summarized in Fig. \ref{fig:phase_diagram}.
For small NNN hopping there is a transition from the QHI 
to the N\'eel antiferromagnet (AF) 
upon increasing $U$ for $\Delta<2t$ as can be seen in Fig. \ref{fig:phase_diagram}(a).
For $\Delta>2t$ the QHI separates the NI at weak from the N\'eel AF at strong $U$.
For the large NNN hopping in Fig. \ref{fig:phase_diagram}(b) we find that the MI is a stripe AF. 
An even more interesting difference compared to the small NNN hopping case is the emergence of a 
$\mathcal{C}=1$ stripe AFQHI in the limit $U\!\sim \! 2\Delta \gg t$.  
We discuss how the compensation of the spin-flip transformation by a lattice translation 
prevents a $\mathcal{C}=1$ N\'eel AFQHI to appear at small NNN hopping. 
We present results for the spectral function in the bulk and at the edges. 
We identify gapless edge states for both spin components in the QHI, and gapless edge states 
for only one spin component in the $\mathcal{C}\!=\!1$ stripe AFQHI.

\section{Method}
Dynamical mean-field theory (DMFT) is a highly successful approach to the problem of strongly correlated 
systems and is exact in the limit of infinite coordination number. For a 
finite coordination number it is an approximation neglecting the momentum dependence 
of the self-energy, or the non-local quantum fluctuations \cite{Kotliar2004,Georges1996,Metzner1989}. 
The $\mathcal{C}=1$ AFQHI phase predicted by DMFT in the Haldane-Hubbard model \cite{Vanhala2016} 
is confirmed by exact diagonalization 
of finite clusters \cite{Vanhala2016} as well as by bold diagrammatic quantum Monte Carlo 
analysis \cite{Tupitsyn2019}. A systematic study of non-local quantum fluctuations in the 
Haldane-Hubbard model \cite{Mertz2019} indicates that a local self-energy can provide an 
appropriate qualitative description of the topological phase diagram; 
the momentum dependence of the self-energy is only needed to map out the precise 
location of the phase boundaries.

We employ the real-space DMFT (RDMFT) approach to qualitatively
analyze the phase diagram of the Hamiltonian Eq. \eqref{4h.eq}.
The RDMFT was first used to study thin film geometries \cite{Potthoff1999}, 
and since then has been extended, for example, to address disordered systems \cite{Song2008,Zheng2019}, 
exotic magnetism \citep{Snoek2008,Irsigler2019a,Orth2013,Hafez-Torbati2019,Valli2018}, 
and topological insulators \citep{Cocks2012,Irsigler2019,Hafez-Torbati2020,Amaricci2018}.
The local self-energy in the DMFT method \cite{Georges1996} becomes position-dependent in the real-space extension, 
allowing for an equal-footing treatment of translationally ordered and disordered systems. 

We use the RDMFT implementation introduced in Ref. \onlinecite{Hafez-Torbati2018}.
We consider $40 \times 40$ lattice sizes with periodic boundary conditions (PBC) in 
both directions unless mentioned otherwise. 
For selected points close to the phase transitions we have checked that 
increasing the systems size to $60 \times 60$ does not change the results.
The temperature is fixed to $T=t/50$, which is much smaller than the energy scales in 
the system and we expect to represent the ground state properties of the model. 
We use exact diagonalization (ED) as the impurity solver \cite{Georges1996,Caffarel1994}. 
Five bath sites are used for the results that we present unless mentioned otherwise.
We have checked that the results for different selected points close to the phase transitions are 
the same as the results obtained using six and seven bath sites.

The Chern number of the interacting system is determined using the topological 
Hamiltonian method \cite{Wang2012}, which relates the Chern number of an interacting 
system to the Chern number of an effective non-interacting model. The method relies 
on the adiabatic deformation of the Green's function such that the single-particle gap 
never closes, leaving the Chern number of the system unchanged.
The effective model, called 
topological Hamiltonian, in the Bloch form reads
\be
{\bs h}_{\rm top}(\vec{k})={\bs h}_0(\vec{k})+{\bs \Sigma}(\vec{k},\omega=0),
\label{eq:htop}
\ee
where ${\bs h}_0(\vec{k})$ describes the non-interacting part of the model 
and ${\bs \Sigma}(\vec{k},\omega)$ is the self-energy. In the DMFT the 
self-energy is local and hence its role in the topological Hamiltonian Eq. \eqref{eq:htop} is 
just to modify the onsite energies \cite{zeroself}. 
One notices that although the topological Hamiltonian method has
some limitations and should be used with care \cite{He2016}, it has been applied successfully 
to similar models \cite{Vanhala2016}.

\begin{figure}[t]
   \begin{center}
   \includegraphics[width=0.53\textwidth,angle=-90]{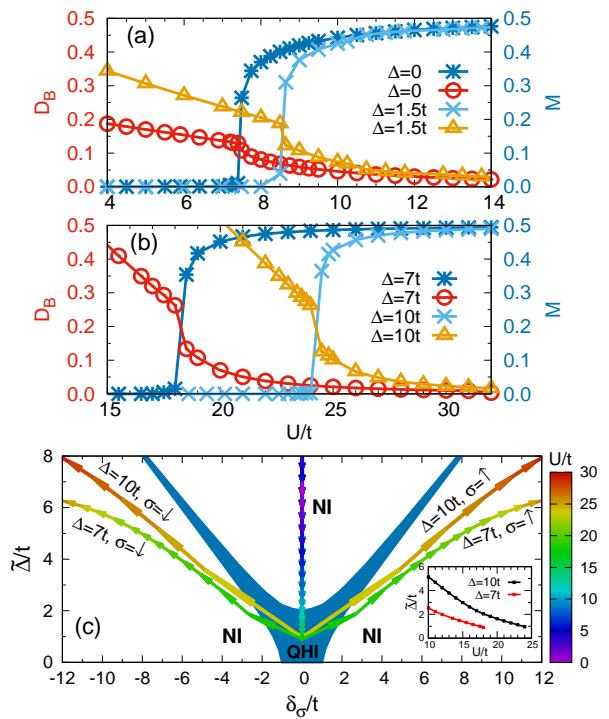}
   \caption{(a,b) The local magnetic moment $M$ and the double occupancy $D_B$ 
   on sublattice $B$  
    plotted versus the Hubbard interaction $U$ for different values of the 
    staggered potential $\Delta$. 
   (c) The evolution of the effective potentials $\tilde{\Delta}$ 
   and $\delta_{\sigma}$ upon increasing $U$ for $\Delta=7t$ and $\Delta=10t$. 
   Here the color indicates the value of $U$ (see the colorbar).
   The shaded area indicates a quantum Hall insulator (QHI) and the white area a normal 
   insulator (NI). The inset shows $\tilde{\Delta}$ versus $U$ in the paramagnetic region 
   where $\delta_\sigma=0$. 
   The results 
   are for the next-nearest-neighbor hopping $t'=0.25t$.}
   \label{Mag_Dou1.fig}
   \end{center}
\end{figure}

\section{Results}
We present results first for the small $t'=0.25t$ and then for the large $t'=t$ NNN hopping. 
We avoid the intermediate values $0.6t \lesssim t' \lesssim 0.8t$ where in the large-$U$ limit 
a quantum spin liquid \cite{Jiang2012,Hu2013,Poilblanc2019} or 
a valence bond crystal \cite{Poilblanc2019,Haghshenas2018,Wang2016,Capriotti2000} is expected, 
which can not be captured within our local self-energy approximation.
One notices that the Hamiltonian Eq. \eqref{4h.eq} in the large-$U$ limit is equivalent, 
up to a weak spatial anisotropy, to the frustrated Heisenberg model with NN and NNN interaction.
For $t'=0.25t$ in Figs. \ref{Mag_Dou1.fig}(a) and \ref{Mag_Dou1.fig}(b) 
we have plotted the local magnetic moment 
$M^\vpdag_{\vecr}=|\langle n_{\vecr,\up}-n_{\vecr,\down} \rangle|/2$ and the double 
occupancy $D^\vpdag_{\vecr}=\langle n_{\vecr,\up} n_{\vecr,\down}\rangle$ versus the Hubbard $U$ 
for different values of the staggered potential $\Delta$. The local 
moment is position-independent, $M_{\vecr}=:M$, and we have plotted the double 
occupancy on sublattice $B$, shown as $D_{\!B}^\vpdag$.
One can identify a transition between a paramagnetic and a magnetic phase, which is 
shifted to larger values of $U$ as $\Delta$ is increased. 
The paramagnetic phase can be a NI or a QHI, depending on the value of the Chern number $\mathcal{C}$. 
The magnetic phase is a N\'eel AF denoted schematically in Fig. \ref{4H.fig}(b).

There are four sites in the 
unit cell labeled as $A_1$, $A_2$, $B_1$, and $B_2$ in Fig. \ref{4H.fig}(b). 
The topological Hamiltonian, in the second quantization form, up to an irrelevant constant 
can be written as
\be
H_{\rm top}=H_{t}+\sum_{\vecr,\sigma}\left( \tilde{\Delta}(-1)^{x}+\delta_{\sigma}(-1)^{x+y}\right) 
{n}^\vpdag_{\vecr,\sigma}
 \label{eq:htop_w}
\ee
where $H_t$ is the hopping term Eq. \eqref{eq:hopping} and the effective potentials $\tilde{\Delta}$ 
and $\delta_\sigma$, in the spirit of Refs. \cite{Hafez-Torbati2020,Amaricci2015,Crippa2020}, are given by 
\begin{subequations}
\label{eq:eff_pot_w}
 \begin{align}
 \label{eq:eff_pot_w:a}
  \tilde{\Delta} & \!=\! \Delta \!+\!\frac{1}{4}\left( \Sigma_{A_1}^{\sigma}(0) \!+\! \Sigma_{A_2}^{\sigma}(0) 
  \!-\! \Sigma_{B_1}^{\sigma}(0) \!-\! \Sigma_{B_2}^{\sigma}(0)\right), \\ 
  \delta_\sigma & \!=\!  \frac{1}{4} 
  \left( 
  \Sigma_{A_1}^{\sigma}(0) \!-\! \Sigma_{A_2}^{\sigma}(0) \!-\! \Sigma_{B_1}^{\sigma}(0) \!+\! \Sigma_{B_2}^{\sigma}(0)
  \right),
 \end{align}
\end{subequations}
where $\Sigma_{X}^{\sigma}(0)$ is the zero-frequency self-energy at the site $X$ with 
spin $\sigma$. 
$\tilde{\Delta}$ is spin-independent and $\delta_\up=-\delta_\down$, see Appendix \ref{app:sec:th}.

\begin{figure}[t]
   \begin{center}
   \includegraphics[width=0.37\textwidth,angle=-90]{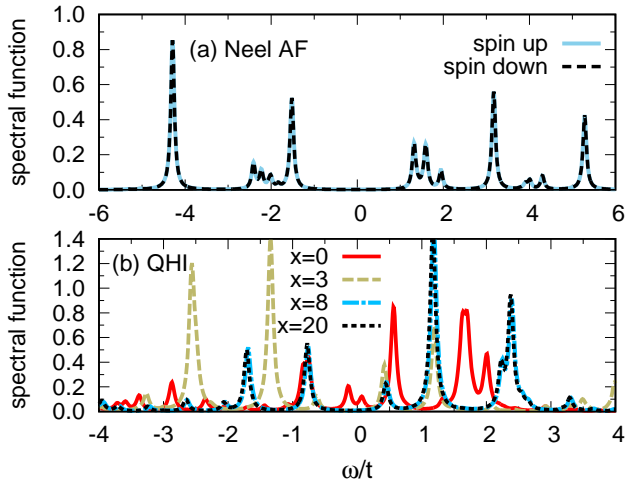}
   \caption{(a) The bulk spectral function averaged over the sites in the unit cell for up and down 
   spin in the N\'eel antiferromagnet (AF) 
   with $\Delta=7t$ and $U=20t$. (b) The spectral function plotted for different values of $x$ 
   in the quantum Hall insulator (QHI) with  $\Delta=7t$ and $U=15t$ obtained using a cylindrical geometry 
   with edges at $x=0$ and $x=40$. The results are for the next-nearest-neighbor hopping 
   $t'=0.25t$.}
   \label{SF1.fig}
   \end{center}
\end{figure}

The evolution of the effective potentials $\tilde{\Delta}$ and $\delta_\sigma$ upon increasing $U$ 
for $\Delta=7t$ and $\Delta=10t$ is displayed in Fig. \ref{Mag_Dou1.fig}(c). The shaded area in  
this figure indicates a QHI and the white area a NI 
with $\tilde{\Delta}$ and $\delta_\sigma$ treated as independent parameters.
Upon increasing $U$ the effective potential $\tilde{\Delta}$ 
is renormalized \cite{Hafez-Torbati2020,Amaricci2015} 
and the system enters the QHI for $\tilde{\Delta}<2t$. This is evident from 
the inset in Fig. \ref{Mag_Dou1.fig}(c) displaying 
$\tilde{\Delta}$ versus $U$ in the paramagnetic region where $\delta_\sigma=0$.
Upon entering the magnetic phase the effective potential $\delta_\sigma$
becomes finite and both spin components fall out of the QHI region \cite{neelc2}. 
This demonstrates that the N\'eel AF is topologically trivial.

It is apparent from Eq. \eqref{eq:htop_w} that the two spin components are always 
in the same topological state due to $\delta_{\up}=-\delta_{\down}$. This makes 
the emergence of a $\mathcal{C}=1$ N\'eel AF impossible. 
This can also be understood from the symmetry of the phase, without considering the 
topological Hamiltonian Eq. \eqref{eq:htop_w}. In the N\'eel AF illustrated in Fig. \ref{4H.fig}(b) 
the effect of the spin-flip transformation can be compensated by a lattice translation, 
i.e., by a shift by one lattice site along $\hat{y}$-direction. 
This suggests that spin up and spin down fermion dispersions will differ at most 
by a shift in momentum space.
This is confirmed in Fig. \ref{SF1.fig}(a) which shows an equal spectral function for up and down spin.
The spectral function is plotted for $-6t \leq \omega \leq+6t$.
The spectral function in Fig. \ref{SF1.fig}(a) is for $\Delta=7t$ and $U=20t$ in 
the N\'eel AF and is averaged over the sites in the unit cell.
The spectral function at position $\vecr$ with spin $\sigma$ is defined from the 
local Green's function as 
$A_{\vec{r},\sigma}(\omega)=(-1/\pi){\rm Im} G_{\vec{r},\sigma}(\omega +i\eta)$ where 
$\eta$ is a broadening factor fixed to $0.05t$ in our computations. 
The  single-particle gap equal for up and down spins prevents a $\mathcal{C}=1$ N\'eel AF from emerging.
The spectral function for $\Delta=7t$ and $U=15t$ and different values of $x$ on 
a $41\times40$ lattice with open boundary conditions (OBC) along $\hat{x}$ and PBC along $\hat{y}$ 
is displayed in Fig. \ref{SF1.fig}(b). 
The edges are defined at $x=0$ and $x=40$ and the lattice is symmetric with respect to the center $x=20$. 
Six bath sites are used in the impurity problem.
There are gapless excitations at the edge which quickly disappear upon approaching the bulk, 
consistent with the  topological Hamiltonian prediction on a QHI phase. 

We consider now the large NNN hopping $t'=t$. The MI phase in this case is a stripe AF. 
The antiferromagnetic ordering is formed along $\hat{x}$ and the ferromagnetic ordering  
along $\hat{y}$, see Fig. \ref{4H.fig}(c), due to the spatial anisotropy induced 
by the staggered potential $\Delta$. 
There are two sites in the unit cell and the topological Hamiltonian for $t'=t$
can be expressed, up to an irrelevant constant, as 
\be
H_{\rm top}=H_t+
\sum_{\vecr,\sigma} \tilde{\Delta}_\sigma (-1)^{x} {n}_{\vecr,\sigma},
\label{eq:htop_s}
\ee
with the effective potential
\be
\tilde{\Delta}_\sigma = \Delta +\frac{1}{2} 
\left( \Sigma_{A}^{\sigma}(0) \!-\! \Sigma_{B}^{\sigma}(0) \right),
\ee
The spin-dependence of this effective potential allows different spin components 
to fall in different topological regions and consequently a $\mathcal{C}=1$ AFQHI to 
emerge. The spin component $\sigma$ is in the quantum Hall state if $|\tilde{\Delta}_\sigma|<2t$
and in the normal state if $|\tilde{\Delta}_\sigma|>2t$.

\begin{figure}[t]
   \begin{center}
   \includegraphics[width=0.52\textwidth,angle=-90]{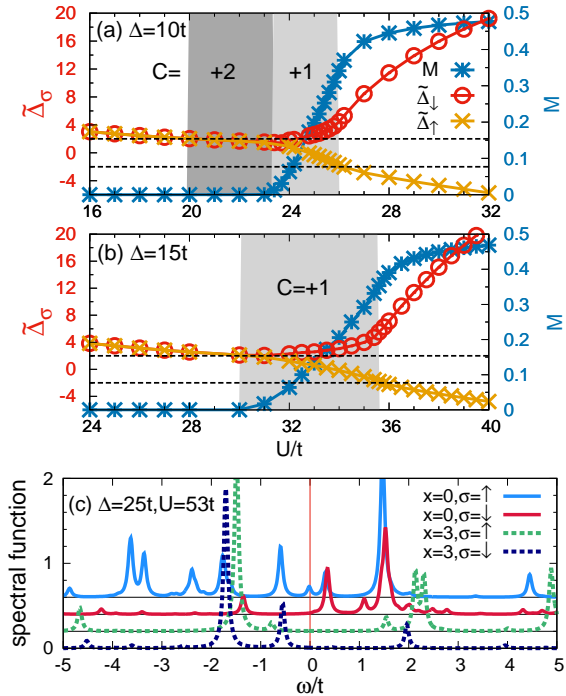}
   \caption{The local magnetic moment $M$ and the effective potential $\tilde{\Delta}_\sigma$ 
   plotted versus the Hubbard interaction $U$ for $\Delta=10t$ (a) and $\Delta=15t$ (b). A shaded area indicates 
   a phase with a finite Chern number $\mathcal{C}$. (c) The edge spectral functions for up and down spin in 
   the $\mathcal{C}=1$ stripe antiferromagnetic quantum Hall insulator with $\Delta=25t$ and $U=53t$, obtained 
   using a cylindrical geometry with edges at $x=0$ and $x=40$. The shift of the spectral function along 
   the vertical axis is for clarity. The results are for the next-nearest-neighbor 
   hopping $t'=t$.}
   \label{M_DT1.fig}
   \end{center}
\end{figure}

In Figs. \ref{M_DT1.fig}(a) and \ref{M_DT1.fig}(b) we have plotted the local magnetic moment $M$
and the effective potential $\tilde{\Delta}_\sigma$ versus $U$ for $\Delta=10t$ (a) and 
$\Delta=15t$ (b). The dashed lines at $\tilde{\Delta}_\sigma=2t$ and 
$\tilde{\Delta}_\sigma=-2t$ specify the 
borders of the topological region $|\tilde{\Delta}_\sigma|<2t$.
A shaded area indicates a phase with a finite Chern number $\mathcal{C}$.
One can see from Fig. \ref{M_DT1.fig}(a) that upon increasing $U$ 
the effective potential $\tilde{\Delta}_\sigma$ drops below $2t$ at $U\simeq 20t$ and 
a transition from a NI to a QHI takes place. For $U \gtrsim 23t$ the local magnetic moment becomes 
finite and $\tilde{\Delta}_\sigma$ becomes spin-dependent. One spin component, spin down in the 
figure, almost immediately leaves the topological region while the other spin component 
remains topological up to $U\simeq 26t$ \cite{stripec2}. 
This leads to a $\mathcal{C}=1$ stripe AFQHI 
phase for $23t \lesssim U\lesssim 26t$. Beyond $U\simeq 26t$ the system is a 
(topologically trivial) stripe AF. 
One can see from Fig. \ref{M_DT1.fig}(b) that upon increasing $\Delta$ to $15t$ the 
QHI phase disappears and there is only the $\mathcal{C}=1$ stripe AFQHI between 
the NI and the stripe AF.

In the stripe AF displayed in Fig. \ref{4H.fig}(c) the effect of the spin-flip 
transformation, unlike the N\'eel AF, can not be compensated by a lattice translation. 
This leads to a spin-dependent spectral function, see Appendix \ref{app:sec:sf}. 
This allows up and down spin components to change their Chern numbers at different 
transition points and the $\mathcal{C}=1$ stripe AFQHI to emerge.

In Fig. \ref{M_DT1.fig}(c) we have plotted the spectral function near the edge $x=0$
of a $41\!\times\!40$ cylindrical geometry 
with $\Delta=25t$ and $U=53t$, where the system is expected to 
be a $\mathcal{C}=1$ stripe AFQHI according to the topological Hamiltonian. 
The shift of the spectral function along the vertical axis is for clarity.
Six bath sites are used in the impurity problem.
There are contributions out of the plotted region $-5t\! \leq \!\omega \! \leq \!+5t$ which mainly 
belong to the spin down spectral function.
Edge excitations in an interacting QHI have been discussed 
using ED on finite clusters \cite{Varney2010} 
and using RDMFT with ED \cite{Hafez-Torbati2020} and with the quantum Monte Carlo \cite{Cocks2012} 
impurity solver. 
We are not aware of a 
study of edge excitations in an interacting $\mathcal{C}\!=\!1$ AFQHI. 
Although our results in Fig. \ref{M_DT1.fig}(c) are obtained using a finite number of 
bath sites and indicate only the qualitative shape of the spectral function, they can 
still capture the main expected feature that edge excitations are gapless for one spin 
component and gapped for the other.  
The edge excitations in optical lattices can be investigated by introducing a Hofstadter 
interface \cite{Irsigler2019}.

\section{Summary}

To summarize, we compare in Fig. \ref{fig:phase_diagram} the $U$-$\Delta$ phase diagram 
of the model Eq. \eqref{4h.eq} for small $t'=0.25t$ (a) and large $t'=t$ (b) NNN hopping.
Apart from the type of magnetic order,
there is a fundamental difference between the two phase diagrams: In Fig. \ref{fig:phase_diagram}(b)
there exists an intermediate $\mathcal{C}=1$ stripe AFQHI while in Fig. \ref{fig:phase_diagram}(a) 
never a $\mathcal{C}=1$ N\'eel AFQHI appears. 
The absence of the AFQHI in the latter case stems from the fact that the effect of the 
spin-flip transformation can be compensated by a space group operation.

We notice that 
our conclusion on the possible existence of a $\mathcal{C}=1$ AFQHI is based on the symmetry 
of the phase and not the details of the model studied in this paper.
For example, replacing the staggered potential along $\hat{x}$ in Eq. \eqref{4h.eq} with 
the staggered potential $H_{\Delta}=\sum_{\vecr, \sigma}\Delta(-1)^{x+y}n_{\vecr,\sigma}$ 
changing along both $\hat{x}$ and $\hat{y}$ directions 
would lead to the opposite situation, i.e., would allow a 
$\mathcal{C}=1$ N\'eel and prevent a $\mathcal{C}=1$ stripe AFQHI. 
Our conclusion is consistent with the realization of the $\mathcal{C}=1$ AFQHI in the
Haldane-Hubbard model \cite{He2011,Vanhala2016}. 
Our results can be used as a guideline for future experiments, 
especially in optical lattices, searching for AFQHI phases.

\section*{Acknowledgements}
We would like to thank Amir A. Ahmad, B. Irsigler, G.S. Uhrig, and J.-H. Zheng for useful discussions.
We are indebted to A. Amaricci for reading the initial version of the manuscript and providing 
helpful comments.
This work  was  supported  by  the  Deutsche  Forschungsgemeinschaft 
(DFG,  German  Research  Foundation)  
via  Research  Unit  FOR  2414  under  Project No.  277974659 (M.H.-T. and W.H.). 
This  work was also  supported  by the  DFG  via  the  high  performance computing center LOEWE-CSC.
This study has also been supported financially by the German Research Foundation (DFG) and the Russian Foundation 
for Basic Research (RFBR) in the International 
Collaborative Research Centre TRR 160, project B8 (M.H.-T.).

\appendix

\section{Topological Hamiltonian for small next-nearest-neighbor hoppings}
\label{app:sec:th}

In this section we derive the topological Hamiltonian Eq. \eqref{eq:htop_w}, which is
valid for small next-nearest-neighbor (NNN) hoppings, i.e., for
the case that in the large-$U$ limit the system exhibits a N\'eel antiferromagnet (AF).
In general, there are four sites in the unit cell as shown in Fig. \ref{4H.fig}(b).
A local self-energy in Eq. \eqref{eq:htop} leaves the hopping part
of the non-interacting Hamiltonian unchanged and modifies only the onsite energies.
One finds
\begin{subequations}
 \label{onsite_top0.eq}
\begin{align}
 \varepsilon_{A_{1},\sigma}^\vpdag&=+\Delta+ \Sigma^{\sigma}_{A_{1}}(0),\\
 \varepsilon_{A_{2},\sigma}^\vpdag&=+\Delta+ \Sigma^{\sigma}_{A_{2}}(0),\\
 \varepsilon_{B_{1},\sigma}^\vpdag&=-\Delta+ \Sigma^{\sigma}_{B_{1}}(0),\\
 \varepsilon_{B_{2},\sigma}^\vpdag&=-\Delta+ \Sigma^{\sigma}_{B_{2}}(0),
\end{align}
\end{subequations}
where $\varepsilon_{X,\sigma}^\vpdag$ represents the onsite energy of the topological 
Hamiltonian at the position $X$ for the spin component $\sigma$. As one can see from 
Fig. \ref{4H.fig}(b) the N\'eel AF is invariant under a spin-flip 
transformation followed by a one-site lattice translation along $\hat{y}$ direction. 
This implies the symmetry relation
\begin{subequations}
 \label{eq:symm}
\begin{align}
 \label{eq:symm1}
 \Sigma^{\sigma}_{A_{1}}(\omega)=\Sigma^{\bar{\sigma}}_{A_{2}}(\omega)
 \quad , \quad
 \Sigma^{\sigma}_{B_{1}}(\omega)=\Sigma^{\bar{\sigma}}_{B_{2}}(\omega),
\end{align}
where $\bar{\sigma}$ indicates the opposite direction of $\sigma$. 
There is the second symmetry relation
\begin{align}
 \label{eq:symm2}
 \Sigma^{\sigma}_{A_{1}}(0)-\Sigma^{\sigma}_{A_{2}}(0)
=
 \Sigma^{\sigma}_{B_{2}}(0)-\Sigma^{\sigma}_{B_{1}}(0),
\end{align}
\end{subequations}
which we found from our data and is valid only at zero frequency. 
Eq. \eqref{onsite_top0.eq} can be rewritten as
\begin{subequations}
 \label{onsite_top1.eq}
\begin{align}
 \varepsilon_{A_{1},\sigma}^\vpdag&=+\Delta+ \Sigma^{\vpdag}_{A_{+}}+\Sigma^{\sigma}_{A_{-}},\\
 \varepsilon_{A_{2},\sigma}^\vpdag&=+\Delta+ \Sigma^{\vpdag}_{A_{+}}-\Sigma^{\sigma}_{A_{-}},\\
 \varepsilon_{B_{1},\sigma}^\vpdag&=-\Delta+ \Sigma^{\vpdag}_{B_{+}}+\Sigma^{\sigma}_{B_{-}},\\
 \varepsilon_{B_{2},\sigma}^\vpdag&=-\Delta+ \Sigma^{\vpdag}_{B_{+}}-\Sigma^{\sigma}_{B_{-}},
\end{align}
\end{subequations}
where we have defined 
\begin{subequations}
\begin{align}
\Sigma^{\vpdag}_{A_{+}}:=\frac{1}{2}\left(\Sigma^{\sigma}_{A_{1}}(0)+\Sigma^{\sigma}_{A_{2}}(0)\right), \\
\Sigma^{\sigma}_{A_{-}}:=\frac{1}{2}\left(\Sigma^{\sigma}_{A_{1}}(0)-\Sigma^{\sigma}_{A_{2}}(0)\right),
\end{align}
\end{subequations}
and similarly for $\Sigma^{\vpdag}_{B_{+}}$ and $\Sigma^{\sigma}_{B_{-}}$. 
$\Sigma^{\vpdag}_{A_{+}}$ and $\Sigma^{\vpdag}_{B_{+}}$ are independent from $\sigma$, 
and $\Sigma^{\sigma}_{A_{-}}=-\Sigma^{\bar{\sigma}}_{A_{-}}$ and $\Sigma^{\sigma}_{B_{-}}=-\Sigma^{\bar{\sigma}}_{B_{-}}$
due to the symmetry relation Eq. \eqref{eq:symm1}. The symmetry relation Eq. \eqref{eq:symm2} 
implies $\Sigma^{\sigma}_{A_{-}}=-\Sigma^{\sigma}_{B_{-}}$. By some straightforward manipulation of 
Eq. \eqref{onsite_top1.eq} we get
\begin{subequations}
 \label{onsite_top2.eq}
\begin{align}
 \varepsilon_{A_{1},\sigma}^\vpdag&=C+\tilde{\Delta}+ \delta_{\sigma}\\
 \varepsilon_{A_{2},\sigma}^\vpdag&=C+\tilde{\Delta}- \delta_{\sigma}\\
 \varepsilon_{B_{1},\sigma}^\vpdag&=C-\tilde{\Delta}- \delta_{\sigma}\\
 \varepsilon_{B_{2},\sigma}^\vpdag&=C-\tilde{\Delta}+ \delta_{\sigma}
\end{align}
\end{subequations}
where we have defined the common constant $C:=(\Sigma^{\vpdag}_{A_{+}}+\Sigma^{\vpdag}_{B_{+}})/2$ 
and the effective potentials 
\begin{subequations}
 \label{eq:eff_pot}
\begin{align}
\tilde{\Delta}&:=\Delta+\frac{1}{2}\left( \Sigma^{\vpdag}_{A_{+}}-\Sigma^{\vpdag}_{B_{+}} \right), \\
\delta^\vpdag_{\sigma}&:=\frac{1}{2}\left( \Sigma^{\sigma}_{A_{-}}-\Sigma^{\sigma}_{B_{-}} \right).
\end{align}
\end{subequations}
One notices that $\tilde{\Delta}$ is independent from $\sigma$ and 
$\delta^\vpdag_{\sigma}=-\delta^\vpdag_{\bar{\sigma}}$ due to the symmetry relations Eq. \eqref{eq:symm}.
This completes the derivation of Eq. \eqref{eq:htop_w} with the effective potentials 
Eq. \eqref{eq:eff_pot_w}.

\section{Spectral function in the stripe antiferromagnetic phase}
\label{app:sec:sf}

\begin{figure}[t]
   \begin{center}
   \includegraphics[width=0.256\textwidth,angle=-90]{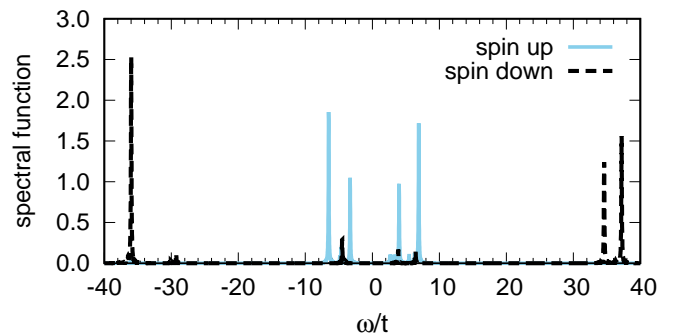}
   \caption{The spectral function in the stripe antiferromagnetic phase for up and down spins 
   plotted versus the frequency $\omega$. The results are for the staggered potential $\Delta=15t$, the Hubbard
   interaction $U=40t$, and the next-nearest-neighbor hopping $t'=t$.}
   \label{fig:spectral_stripe}
   \end{center}
\end{figure}

In Fig. \ref{fig:spectral_stripe} we have plotted the bulk spectral function averaged over the sites 
in the unit cell in the stripe antiferromagnet (AF) for up and down spins. 
The results are for the staggered potential $\Delta=15t$, the Hubbard
interaction $U=40t$, and the next-nearest-neighbor hopping $t'=t$. 
In contrast to the spectral function in the N\'eel AF in Fig. \ref{SF1.fig}(a), the 
spectral function in the stripe AF depends on spin. This is due to the fact that the effect of the 
spin-flip transformation can not be compensated by a space group operation in the stripe AF, 
see Fig. \ref{4H.fig} (c).

\section*{references}

\begin{thebibliography}{74}%
\makeatletter
\providecommand \@ifxundefined [1]{%
 \@ifx{#1\undefined}
}%
\providecommand \@ifnum [1]{%
 \ifnum #1\expandafter \@firstoftwo
 \else \expandafter \@secondoftwo
 \fi
}%
\providecommand \@ifx [1]{%
 \ifx #1\expandafter \@firstoftwo
 \else \expandafter \@secondoftwo
 \fi
}%
\providecommand \natexlab [1]{#1}%
\providecommand \enquote  [1]{``#1''}%
\providecommand \bibnamefont  [1]{#1}%
\providecommand \bibfnamefont [1]{#1}%
\providecommand \citenamefont [1]{#1}%
\providecommand \href@noop [0]{\@secondoftwo}%
\providecommand \href [0]{\begingroup \@sanitize@url \@href}%
\providecommand \@href[1]{\@@startlink{#1}\@@href}%
\providecommand \@@href[1]{\endgroup#1\@@endlink}%
\providecommand \@sanitize@url [0]{\catcode `\\12\catcode `\$12\catcode
  `\&12\catcode `\#12\catcode `\^12\catcode `\_12\catcode `\%12\relax}%
\providecommand \@@startlink[1]{}%
\providecommand \@@endlink[0]{}%
\providecommand \url  [0]{\begingroup\@sanitize@url \@url }%
\providecommand \@url [1]{\endgroup\@href {#1}{\urlprefix }}%
\providecommand \urlprefix  [0]{URL }%
\providecommand \Eprint [0]{\href }%
\providecommand \doibase [0]{http://dx.doi.org/}%
\providecommand \selectlanguage [0]{\@gobble}%
\providecommand \bibinfo  [0]{\@secondoftwo}%
\providecommand \bibfield  [0]{\@secondoftwo}%
\providecommand \translation [1]{[#1]}%
\providecommand \BibitemOpen [0]{}%
\providecommand \bibitemStop [0]{}%
\providecommand \bibitemNoStop [0]{.\EOS\space}%
\providecommand \EOS [0]{\spacefactor3000\relax}%
\providecommand \BibitemShut  [1]{\csname bibitem#1\endcsname}%
\let\auto@bib@innerbib\@empty
\bibitem [{\citenamefont {Tokura}\ \emph {et~al.}(2019)\citenamefont {Tokura},
  \citenamefont {Yasuda},\ and\ \citenamefont {Tsukazaki}}]{Tokura2019}%
  \BibitemOpen
  \bibfield  {author} {\bibinfo {author} {\bibfnamefont {Y.}~\bibnamefont
  {Tokura}}, \bibinfo {author} {\bibfnamefont {K.}~\bibnamefont {Yasuda}}, \
  and\ \bibinfo {author} {\bibfnamefont {A.}~\bibnamefont {Tsukazaki}},\ }\href
  {https://doi.org/10.1038/s42254-018-0011-5} {\bibfield  {journal} {\bibinfo
  {journal} {Nature Reviews Physics}\ }\textbf {\bibinfo {volume} {1}},\
  \bibinfo {pages} {126} (\bibinfo {year} {2019})}\BibitemShut {NoStop}%
\bibitem [{\citenamefont {Zhang}\ \emph {et~al.}(2019)\citenamefont {Zhang},
  \citenamefont {Shi}, \citenamefont {Zhu}, \citenamefont {Xing}, \citenamefont
  {Zhang},\ and\ \citenamefont {Wang}}]{Zhang2019}%
  \BibitemOpen
  \bibfield  {author} {\bibinfo {author} {\bibfnamefont {D.}~\bibnamefont
  {Zhang}}, \bibinfo {author} {\bibfnamefont {M.}~\bibnamefont {Shi}}, \bibinfo
  {author} {\bibfnamefont {T.}~\bibnamefont {Zhu}}, \bibinfo {author}
  {\bibfnamefont {D.}~\bibnamefont {Xing}}, \bibinfo {author} {\bibfnamefont
  {H.}~\bibnamefont {Zhang}}, \ and\ \bibinfo {author} {\bibfnamefont
  {J.}~\bibnamefont {Wang}},\ }\href {\doibase 10.1103/PhysRevLett.122.206401}
  {\bibfield  {journal} {\bibinfo  {journal} {Phys. Rev. Lett.}\ }\textbf
  {\bibinfo {volume} {122}},\ \bibinfo {pages} {206401} (\bibinfo {year}
  {2019})}\BibitemShut {NoStop}%
\bibitem [{\citenamefont {Otrokov}\ \emph
  {et~al.}(2019{\natexlab{a}})\citenamefont {Otrokov}, \citenamefont {Rusinov},
  \citenamefont {Blanco-Rey}, \citenamefont {Hoffmann}, \citenamefont
  {Vyazovskaya}, \citenamefont {Eremeev}, \citenamefont {Ernst}, \citenamefont
  {Echenique}, \citenamefont {Arnau},\ and\ \citenamefont
  {Chulkov}}]{Otrokov2019a}%
  \BibitemOpen
  \bibfield  {author} {\bibinfo {author} {\bibfnamefont {M.~M.}\ \bibnamefont
  {Otrokov}}, \bibinfo {author} {\bibfnamefont {I.~P.}\ \bibnamefont
  {Rusinov}}, \bibinfo {author} {\bibfnamefont {M.}~\bibnamefont {Blanco-Rey}},
  \bibinfo {author} {\bibfnamefont {M.}~\bibnamefont {Hoffmann}}, \bibinfo
  {author} {\bibfnamefont {A.~Y.}\ \bibnamefont {Vyazovskaya}}, \bibinfo
  {author} {\bibfnamefont {S.~V.}\ \bibnamefont {Eremeev}}, \bibinfo {author}
  {\bibfnamefont {A.}~\bibnamefont {Ernst}}, \bibinfo {author} {\bibfnamefont
  {P.~M.}\ \bibnamefont {Echenique}}, \bibinfo {author} {\bibfnamefont
  {A.}~\bibnamefont {Arnau}}, \ and\ \bibinfo {author} {\bibfnamefont {E.~V.}\
  \bibnamefont {Chulkov}},\ }\href {\doibase 10.1103/PhysRevLett.122.107202}
  {\bibfield  {journal} {\bibinfo  {journal} {Phys. Rev. Lett.}\ }\textbf
  {\bibinfo {volume} {122}},\ \bibinfo {pages} {107202} (\bibinfo {year}
  {2019}{\natexlab{a}})}\BibitemShut {NoStop}%
\bibitem [{\citenamefont {Li}\ \emph {et~al.}(2019{\natexlab{a}})\citenamefont
  {Li}, \citenamefont {Li}, \citenamefont {Du}, \citenamefont {Wang},
  \citenamefont {Gu}, \citenamefont {Zhang}, \citenamefont {He}, \citenamefont
  {Duan},\ and\ \citenamefont {Xu}}]{Li2019a}%
  \BibitemOpen
  \bibfield  {author} {\bibinfo {author} {\bibfnamefont {J.}~\bibnamefont
  {Li}}, \bibinfo {author} {\bibfnamefont {Y.}~\bibnamefont {Li}}, \bibinfo
  {author} {\bibfnamefont {S.}~\bibnamefont {Du}}, \bibinfo {author}
  {\bibfnamefont {Z.}~\bibnamefont {Wang}}, \bibinfo {author} {\bibfnamefont
  {B.-L.}\ \bibnamefont {Gu}}, \bibinfo {author} {\bibfnamefont {S.-C.}\
  \bibnamefont {Zhang}}, \bibinfo {author} {\bibfnamefont {K.}~\bibnamefont
  {He}}, \bibinfo {author} {\bibfnamefont {W.}~\bibnamefont {Duan}}, \ and\
  \bibinfo {author} {\bibfnamefont {Y.}~\bibnamefont {Xu}},\ }\href {\doibase
  10.1126/sciadv.aaw5685} {\bibfield  {journal} {\bibinfo  {journal} {Science
  Advances}\ }\textbf {\bibinfo {volume} {5}},\ \bibinfo {pages} {eaaw5685}
  (\bibinfo {year} {2019}{\natexlab{a}})}\BibitemShut {NoStop}%
\bibitem [{\citenamefont {Li}\ \emph {et~al.}(2019{\natexlab{b}})\citenamefont
  {Li}, \citenamefont {Gao}, \citenamefont {Duan}, \citenamefont {Xu},
  \citenamefont {Zhu}, \citenamefont {Tian}, \citenamefont {Gao}, \citenamefont
  {Fan}, \citenamefont {Rao}, \citenamefont {Huang}, \citenamefont {Li},
  \citenamefont {Yan}, \citenamefont {Liu}, \citenamefont {Liu}, \citenamefont
  {Huang}, \citenamefont {Li}, \citenamefont {Liu}, \citenamefont {Zhang},
  \citenamefont {Zhang}, \citenamefont {Kondo}, \citenamefont {Shin},
  \citenamefont {Lei}, \citenamefont {Shi}, \citenamefont {Zhang},
  \citenamefont {Weng}, \citenamefont {Qian},\ and\ \citenamefont
  {Ding}}]{Li2019}%
  \BibitemOpen
  \bibfield  {author} {\bibinfo {author} {\bibfnamefont {H.}~\bibnamefont
  {Li}}, \bibinfo {author} {\bibfnamefont {S.-Y.}\ \bibnamefont {Gao}},
  \bibinfo {author} {\bibfnamefont {S.-F.}\ \bibnamefont {Duan}}, \bibinfo
  {author} {\bibfnamefont {Y.-F.}\ \bibnamefont {Xu}}, \bibinfo {author}
  {\bibfnamefont {K.-J.}\ \bibnamefont {Zhu}}, \bibinfo {author} {\bibfnamefont
  {S.-J.}\ \bibnamefont {Tian}}, \bibinfo {author} {\bibfnamefont {J.-C.}\
  \bibnamefont {Gao}}, \bibinfo {author} {\bibfnamefont {W.-H.}\ \bibnamefont
  {Fan}}, \bibinfo {author} {\bibfnamefont {Z.-C.}\ \bibnamefont {Rao}},
  \bibinfo {author} {\bibfnamefont {J.-R.}\ \bibnamefont {Huang}}, \bibinfo
  {author} {\bibfnamefont {J.-J.}\ \bibnamefont {Li}}, \bibinfo {author}
  {\bibfnamefont {D.-Y.}\ \bibnamefont {Yan}}, \bibinfo {author} {\bibfnamefont
  {Z.-T.}\ \bibnamefont {Liu}}, \bibinfo {author} {\bibfnamefont {W.-L.}\
  \bibnamefont {Liu}}, \bibinfo {author} {\bibfnamefont {Y.-B.}\ \bibnamefont
  {Huang}}, \bibinfo {author} {\bibfnamefont {Y.-L.}\ \bibnamefont {Li}},
  \bibinfo {author} {\bibfnamefont {Y.}~\bibnamefont {Liu}}, \bibinfo {author}
  {\bibfnamefont {G.-B.}\ \bibnamefont {Zhang}}, \bibinfo {author}
  {\bibfnamefont {P.}~\bibnamefont {Zhang}}, \bibinfo {author} {\bibfnamefont
  {T.}~\bibnamefont {Kondo}}, \bibinfo {author} {\bibfnamefont
  {S.}~\bibnamefont {Shin}}, \bibinfo {author} {\bibfnamefont {H.-C.}\
  \bibnamefont {Lei}}, \bibinfo {author} {\bibfnamefont {Y.-G.}\ \bibnamefont
  {Shi}}, \bibinfo {author} {\bibfnamefont {W.-T.}\ \bibnamefont {Zhang}},
  \bibinfo {author} {\bibfnamefont {H.-M.}\ \bibnamefont {Weng}}, \bibinfo
  {author} {\bibfnamefont {T.}~\bibnamefont {Qian}}, \ and\ \bibinfo {author}
  {\bibfnamefont {H.}~\bibnamefont {Ding}},\ }\href {\doibase
  10.1103/PhysRevX.9.041039} {\bibfield  {journal} {\bibinfo  {journal}
  {Physical Review X}\ }\textbf {\bibinfo {volume} {9}},\ \bibinfo {pages}
  {041039} (\bibinfo {year} {2019}{\natexlab{b}})}\BibitemShut {NoStop}%
\bibitem [{\citenamefont {Otrokov}\ \emph
  {et~al.}(2019{\natexlab{b}})\citenamefont {Otrokov}, \citenamefont
  {Klimovskikh}, \citenamefont {Bentmann}, \citenamefont {Estyunin},
  \citenamefont {Zeugner}, \citenamefont {Aliev}, \citenamefont {Gaß},
  \citenamefont {Wolter}, \citenamefont {Koroleva}, \citenamefont {Shikin},
  \citenamefont {Blanco-Rey}, \citenamefont {Hoffmann}, \citenamefont
  {Rusinov}, \citenamefont {Vyazovskaya}, \citenamefont {Eremeev},
  \citenamefont {Koroteev}, \citenamefont {Kuznetsov}, \citenamefont {Freyse},
  \citenamefont {Sánchez-Barriga}, \citenamefont {Amiraslanov}, \citenamefont
  {Babanly}, \citenamefont {Mamedov}, \citenamefont {Abdullayev}, \citenamefont
  {Zverev}, \citenamefont {Alfonsov}, \citenamefont {Kataev}, \citenamefont
  {Büchner}, \citenamefont {Schwier}, \citenamefont {Kumar}, \citenamefont
  {Kimura}, \citenamefont {Petaccia}, \citenamefont {Di~Santo}, \citenamefont
  {Vidal}, \citenamefont {Schatz}, \citenamefont {Kißner}, \citenamefont
  {Ünzelmann}, \citenamefont {Min}, \citenamefont {Moser}, \citenamefont
  {Peixoto}, \citenamefont {Reinert}, \citenamefont {Ernst}, \citenamefont
  {Echenique}, \citenamefont {Isaeva},\ and\ \citenamefont
  {Chulkov}}]{Otrokov2019}%
  \BibitemOpen
  \bibfield  {author} {\bibinfo {author} {\bibfnamefont {M.~M.}\ \bibnamefont
  {Otrokov}}, \bibinfo {author} {\bibfnamefont {I.~I.}\ \bibnamefont
  {Klimovskikh}}, \bibinfo {author} {\bibfnamefont {H.}~\bibnamefont
  {Bentmann}}, \bibinfo {author} {\bibfnamefont {D.}~\bibnamefont {Estyunin}},
  \bibinfo {author} {\bibfnamefont {A.}~\bibnamefont {Zeugner}}, \bibinfo
  {author} {\bibfnamefont {Z.~S.}\ \bibnamefont {Aliev}}, \bibinfo {author}
  {\bibfnamefont {S.}~\bibnamefont {Gaß}}, \bibinfo {author} {\bibfnamefont
  {A.~U.~B.}\ \bibnamefont {Wolter}}, \bibinfo {author} {\bibfnamefont {A.~V.}\
  \bibnamefont {Koroleva}}, \bibinfo {author} {\bibfnamefont {A.~M.}\
  \bibnamefont {Shikin}}, \bibinfo {author} {\bibfnamefont {M.}~\bibnamefont
  {Blanco-Rey}}, \bibinfo {author} {\bibfnamefont {M.}~\bibnamefont
  {Hoffmann}}, \bibinfo {author} {\bibfnamefont {I.~P.}\ \bibnamefont
  {Rusinov}}, \bibinfo {author} {\bibfnamefont {A.~Y.}\ \bibnamefont
  {Vyazovskaya}}, \bibinfo {author} {\bibfnamefont {S.~V.}\ \bibnamefont
  {Eremeev}}, \bibinfo {author} {\bibfnamefont {Y.~M.}\ \bibnamefont
  {Koroteev}}, \bibinfo {author} {\bibfnamefont {V.~M.}\ \bibnamefont
  {Kuznetsov}}, \bibinfo {author} {\bibfnamefont {F.}~\bibnamefont {Freyse}},
  \bibinfo {author} {\bibfnamefont {J.}~\bibnamefont {Sánchez-Barriga}},
  \bibinfo {author} {\bibfnamefont {I.~R.}\ \bibnamefont {Amiraslanov}},
  \bibinfo {author} {\bibfnamefont {M.~B.}\ \bibnamefont {Babanly}}, \bibinfo
  {author} {\bibfnamefont {N.~T.}\ \bibnamefont {Mamedov}}, \bibinfo {author}
  {\bibfnamefont {N.~A.}\ \bibnamefont {Abdullayev}}, \bibinfo {author}
  {\bibfnamefont {V.~N.}\ \bibnamefont {Zverev}}, \bibinfo {author}
  {\bibfnamefont {A.}~\bibnamefont {Alfonsov}}, \bibinfo {author}
  {\bibfnamefont {V.}~\bibnamefont {Kataev}}, \bibinfo {author} {\bibfnamefont
  {B.}~\bibnamefont {Büchner}}, \bibinfo {author} {\bibfnamefont {E.~F.}\
  \bibnamefont {Schwier}}, \bibinfo {author} {\bibfnamefont {S.}~\bibnamefont
  {Kumar}}, \bibinfo {author} {\bibfnamefont {A.}~\bibnamefont {Kimura}},
  \bibinfo {author} {\bibfnamefont {L.}~\bibnamefont {Petaccia}}, \bibinfo
  {author} {\bibfnamefont {G.}~\bibnamefont {Di~Santo}}, \bibinfo {author}
  {\bibfnamefont {R.~C.}\ \bibnamefont {Vidal}}, \bibinfo {author}
  {\bibfnamefont {S.}~\bibnamefont {Schatz}}, \bibinfo {author} {\bibfnamefont
  {K.}~\bibnamefont {Kißner}}, \bibinfo {author} {\bibfnamefont
  {M.}~\bibnamefont {Ünzelmann}}, \bibinfo {author} {\bibfnamefont {C.~H.}\
  \bibnamefont {Min}}, \bibinfo {author} {\bibfnamefont {S.}~\bibnamefont
  {Moser}}, \bibinfo {author} {\bibfnamefont {T.~R.~F.}\ \bibnamefont
  {Peixoto}}, \bibinfo {author} {\bibfnamefont {F.}~\bibnamefont {Reinert}},
  \bibinfo {author} {\bibfnamefont {A.}~\bibnamefont {Ernst}}, \bibinfo
  {author} {\bibfnamefont {P.~M.}\ \bibnamefont {Echenique}}, \bibinfo {author}
  {\bibfnamefont {A.}~\bibnamefont {Isaeva}}, \ and\ \bibinfo {author}
  {\bibfnamefont {E.~V.}\ \bibnamefont {Chulkov}},\ }\href
  {https://doi.org/10.1038/s41586-019-1840-9} {\bibfield  {journal} {\bibinfo
  {journal} {Nature}\ }\textbf {\bibinfo {volume} {576}},\ \bibinfo {pages}
  {416} (\bibinfo {year} {2019}{\natexlab{b}})}\BibitemShut {NoStop}%
\bibitem [{\citenamefont {Cooper}\ \emph {et~al.}(2019)\citenamefont {Cooper},
  \citenamefont {Dalibard},\ and\ \citenamefont {Spielman}}]{Cooper2019}%
  \BibitemOpen
  \bibfield  {author} {\bibinfo {author} {\bibfnamefont {N.~R.}\ \bibnamefont
  {Cooper}}, \bibinfo {author} {\bibfnamefont {J.}~\bibnamefont {Dalibard}}, \
  and\ \bibinfo {author} {\bibfnamefont {I.~B.}\ \bibnamefont {Spielman}},\
  }\href {https://link.aps.org/doi/10.1103/RevModPhys.91.015005} {\bibfield
  {journal} {\bibinfo  {journal} {Rev. Mod. Phys.}\ }\textbf {\bibinfo {volume}
  {91}},\ \bibinfo {pages} {015005} (\bibinfo {year} {2019})}\BibitemShut
  {NoStop}%
\bibitem [{\citenamefont {Aidelsburger}\ \emph {et~al.}(2018)\citenamefont
  {Aidelsburger}, \citenamefont {Nascimbene},\ and\ \citenamefont
  {Goldman}}]{Aidelsburger2018}%
  \BibitemOpen
  \bibfield  {author} {\bibinfo {author} {\bibfnamefont {M.}~\bibnamefont
  {Aidelsburger}}, \bibinfo {author} {\bibfnamefont {S.}~\bibnamefont
  {Nascimbene}}, \ and\ \bibinfo {author} {\bibfnamefont {N.}~\bibnamefont
  {Goldman}},\ }\href {\doibase 10.1016/j.crhy.2018.03.002} {\bibfield
  {journal} {\bibinfo  {journal} {Comptes Rendus Physique}\ }\textbf {\bibinfo
  {volume} {19}},\ \bibinfo {pages} {394–432} (\bibinfo {year} {2018})},\
  \bibinfo {note} {quantum simulation / Simulation quantique}\BibitemShut
  {NoStop}%
\bibitem [{\citenamefont {Mazurenko}\ \emph {et~al.}(2017)\citenamefont
  {Mazurenko}, \citenamefont {Chiu}, \citenamefont {Ji}, \citenamefont
  {Parsons}, \citenamefont {Kanász-Nagy}, \citenamefont {Schmidt},
  \citenamefont {Grusdt}, \citenamefont {Demler}, \citenamefont {Greif},\ and\
  \citenamefont {Greiner}}]{Mazurenko2017}%
  \BibitemOpen
  \bibfield  {author} {\bibinfo {author} {\bibfnamefont {A.}~\bibnamefont
  {Mazurenko}}, \bibinfo {author} {\bibfnamefont {C.~S.}\ \bibnamefont {Chiu}},
  \bibinfo {author} {\bibfnamefont {G.}~\bibnamefont {Ji}}, \bibinfo {author}
  {\bibfnamefont {M.~F.}\ \bibnamefont {Parsons}}, \bibinfo {author}
  {\bibfnamefont {M.}~\bibnamefont {Kanász-Nagy}}, \bibinfo {author}
  {\bibfnamefont {R.}~\bibnamefont {Schmidt}}, \bibinfo {author} {\bibfnamefont
  {F.}~\bibnamefont {Grusdt}}, \bibinfo {author} {\bibfnamefont
  {E.}~\bibnamefont {Demler}}, \bibinfo {author} {\bibfnamefont
  {D.}~\bibnamefont {Greif}}, \ and\ \bibinfo {author} {\bibfnamefont
  {M.}~\bibnamefont {Greiner}},\ }\href {https://doi.org/10.1038/nature22362}
  {\bibfield  {journal} {\bibinfo  {journal} {Nature}\ }\textbf {\bibinfo
  {volume} {545}},\ \bibinfo {pages} {462} (\bibinfo {year}
  {2017})}\BibitemShut {NoStop}%
\bibitem [{\citenamefont {Brown}\ \emph {et~al.}(2017)\citenamefont {Brown},
  \citenamefont {Mitra}, \citenamefont {Guardado-Sanchez}, \citenamefont
  {Schauß}, \citenamefont {Kondov}, \citenamefont {Khatami}, \citenamefont
  {Paiva}, \citenamefont {Trivedi}, \citenamefont {Huse},\ and\ \citenamefont
  {Bakr}}]{Brown2017}%
  \BibitemOpen
  \bibfield  {author} {\bibinfo {author} {\bibfnamefont {P.~T.}\ \bibnamefont
  {Brown}}, \bibinfo {author} {\bibfnamefont {D.}~\bibnamefont {Mitra}},
  \bibinfo {author} {\bibfnamefont {E.}~\bibnamefont {Guardado-Sanchez}},
  \bibinfo {author} {\bibfnamefont {P.}~\bibnamefont {Schauß}}, \bibinfo
  {author} {\bibfnamefont {S.~S.}\ \bibnamefont {Kondov}}, \bibinfo {author}
  {\bibfnamefont {E.}~\bibnamefont {Khatami}}, \bibinfo {author} {\bibfnamefont
  {T.}~\bibnamefont {Paiva}}, \bibinfo {author} {\bibfnamefont
  {N.}~\bibnamefont {Trivedi}}, \bibinfo {author} {\bibfnamefont {D.~A.}\
  \bibnamefont {Huse}}, \ and\ \bibinfo {author} {\bibfnamefont {W.~S.}\
  \bibnamefont {Bakr}},\ }\href {\doibase 10.1126/science.aam7838} {\bibfield
  {journal} {\bibinfo  {journal} {Science}\ }\textbf {\bibinfo {volume}
  {357}},\ \bibinfo {pages} {1385–1388} (\bibinfo {year} {2017})}\BibitemShut
  {NoStop}%
\bibitem [{\citenamefont {Hofstetter}\ and\ \citenamefont
  {Qin}(2018)}]{Hofstetter2018}%
  \BibitemOpen
  \bibfield  {author} {\bibinfo {author} {\bibfnamefont {W.}~\bibnamefont
  {Hofstetter}}\ and\ \bibinfo {author} {\bibfnamefont {T.}~\bibnamefont
  {Qin}},\ }\href {http://stacks.iop.org/0953-4075/51/i=8/a=082001} {\bibfield
  {journal} {\bibinfo  {journal} {Journal of Physics B: Atomic, Molecular and
  Optical Physics}\ }\textbf {\bibinfo {volume} {51}},\ \bibinfo {pages}
  {082001} (\bibinfo {year} {2018})}\BibitemShut {NoStop}%
\bibitem [{\citenamefont {Aidelsburger}\ \emph {et~al.}(2013)\citenamefont
  {Aidelsburger}, \citenamefont {Atala}, \citenamefont {Lohse}, \citenamefont
  {Barreiro}, \citenamefont {Paredes},\ and\ \citenamefont
  {Bloch}}]{Aidelsburger2013}%
  \BibitemOpen
  \bibfield  {author} {\bibinfo {author} {\bibfnamefont {M.}~\bibnamefont
  {Aidelsburger}}, \bibinfo {author} {\bibfnamefont {M.}~\bibnamefont {Atala}},
  \bibinfo {author} {\bibfnamefont {M.}~\bibnamefont {Lohse}}, \bibinfo
  {author} {\bibfnamefont {J.~T.}\ \bibnamefont {Barreiro}}, \bibinfo {author}
  {\bibfnamefont {B.}~\bibnamefont {Paredes}}, \ and\ \bibinfo {author}
  {\bibfnamefont {I.}~\bibnamefont {Bloch}},\ }\href {\doibase
  10.1103/PhysRevLett.111.185301} {\bibfield  {journal} {\bibinfo  {journal}
  {Phys. Rev. Lett.}\ }\textbf {\bibinfo {volume} {111}},\ \bibinfo {pages}
  {185301} (\bibinfo {year} {2013})}\BibitemShut {NoStop}%
\bibitem [{\citenamefont {Miyake}\ \emph {et~al.}(2013)\citenamefont {Miyake},
  \citenamefont {Siviloglou}, \citenamefont {Kennedy}, \citenamefont {Burton},\
  and\ \citenamefont {Ketterle}}]{Miyake2013}%
  \BibitemOpen
  \bibfield  {author} {\bibinfo {author} {\bibfnamefont {H.}~\bibnamefont
  {Miyake}}, \bibinfo {author} {\bibfnamefont {G.~A.}\ \bibnamefont
  {Siviloglou}}, \bibinfo {author} {\bibfnamefont {C.~J.}\ \bibnamefont
  {Kennedy}}, \bibinfo {author} {\bibfnamefont {W.~C.}\ \bibnamefont {Burton}},
  \ and\ \bibinfo {author} {\bibfnamefont {W.}~\bibnamefont {Ketterle}},\
  }\href {\doibase 10.1103/PhysRevLett.111.185302} {\bibfield  {journal}
  {\bibinfo  {journal} {Phys. Rev. Lett.}\ }\textbf {\bibinfo {volume} {111}},\
  \bibinfo {pages} {185302} (\bibinfo {year} {2013})}\BibitemShut {NoStop}%
\bibitem [{\citenamefont {Jotzu}\ \emph {et~al.}(2014)\citenamefont {Jotzu},
  \citenamefont {Messer}, \citenamefont {Desbuquois}, \citenamefont {Lebrat},
  \citenamefont {Uehlinger}, \citenamefont {Greif},\ and\ \citenamefont
  {Esslinger}}]{Jotzu2014}%
  \BibitemOpen
  \bibfield  {author} {\bibinfo {author} {\bibfnamefont {G.}~\bibnamefont
  {Jotzu}}, \bibinfo {author} {\bibfnamefont {M.}~\bibnamefont {Messer}},
  \bibinfo {author} {\bibfnamefont {R.}~\bibnamefont {Desbuquois}}, \bibinfo
  {author} {\bibfnamefont {M.}~\bibnamefont {Lebrat}}, \bibinfo {author}
  {\bibfnamefont {T.}~\bibnamefont {Uehlinger}}, \bibinfo {author}
  {\bibfnamefont {D.}~\bibnamefont {Greif}}, \ and\ \bibinfo {author}
  {\bibfnamefont {T.}~\bibnamefont {Esslinger}},\ }\href
  {https://doi.org/10.1038/nature13915} {\bibfield  {journal} {\bibinfo
  {journal} {Nature}\ }\textbf {\bibinfo {volume} {515}},\ \bibinfo {pages}
  {237} (\bibinfo {year} {2014})}\BibitemShut {NoStop}%
\bibitem [{\citenamefont {Aidelsburger}\ \emph {et~al.}(2014)\citenamefont
  {Aidelsburger}, \citenamefont {Lohse}, \citenamefont {Schweizer},
  \citenamefont {Atala}, \citenamefont {Barreiro}, \citenamefont {Nascimbène},
  \citenamefont {Cooper}, \citenamefont {Bloch},\ and\ \citenamefont
  {Goldman}}]{Aidelsburger2014}%
  \BibitemOpen
  \bibfield  {author} {\bibinfo {author} {\bibfnamefont {M.}~\bibnamefont
  {Aidelsburger}}, \bibinfo {author} {\bibfnamefont {M.}~\bibnamefont {Lohse}},
  \bibinfo {author} {\bibfnamefont {C.}~\bibnamefont {Schweizer}}, \bibinfo
  {author} {\bibfnamefont {M.}~\bibnamefont {Atala}}, \bibinfo {author}
  {\bibfnamefont {J.~T.}\ \bibnamefont {Barreiro}}, \bibinfo {author}
  {\bibfnamefont {S.}~\bibnamefont {Nascimbène}}, \bibinfo {author}
  {\bibfnamefont {N.~R.}\ \bibnamefont {Cooper}}, \bibinfo {author}
  {\bibfnamefont {I.}~\bibnamefont {Bloch}}, \ and\ \bibinfo {author}
  {\bibfnamefont {N.}~\bibnamefont {Goldman}},\ }\href
  {https://doi.org/10.1038/nphys3171} {\bibfield  {journal} {\bibinfo
  {journal} {Nature Physics}\ }\textbf {\bibinfo {volume} {11}},\ \bibinfo
  {pages} {162–} (\bibinfo {year} {2014})}\BibitemShut {NoStop}%
\bibitem [{\citenamefont {Fl\"{a}schner}\ \emph {et~al.}(2016)\citenamefont
  {Fl\"{a}schner}, \citenamefont {Rem}, \citenamefont {Tarnowski},
  \citenamefont {Vogel}, \citenamefont {L\"{u}hmann}, \citenamefont
  {Sengstock},\ and\ \citenamefont {Weitenberg}}]{Flaschner2016}%
  \BibitemOpen
  \bibfield  {author} {\bibinfo {author} {\bibfnamefont {N.}~\bibnamefont
  {Fl\"{a}schner}}, \bibinfo {author} {\bibfnamefont {B.~S.}\ \bibnamefont
  {Rem}}, \bibinfo {author} {\bibfnamefont {M.}~\bibnamefont {Tarnowski}},
  \bibinfo {author} {\bibfnamefont {D.}~\bibnamefont {Vogel}}, \bibinfo
  {author} {\bibfnamefont {D.-S.}\ \bibnamefont {L\"{u}hmann}}, \bibinfo
  {author} {\bibfnamefont {K.}~\bibnamefont {Sengstock}}, \ and\ \bibinfo
  {author} {\bibfnamefont {C.}~\bibnamefont {Weitenberg}},\ }\href
  {http://science.sciencemag.org/content/352/6289/1091.abstract} {\bibfield
  {journal} {\bibinfo  {journal} {Science}\ }\textbf {\bibinfo {volume}
  {352}},\ \bibinfo {pages} {1091} (\bibinfo {year} {2016})}\BibitemShut
  {NoStop}%
\bibitem [{\citenamefont {Bloch}\ \emph {et~al.}(2008)\citenamefont {Bloch},
  \citenamefont {Dalibard},\ and\ \citenamefont {Zwerger}}]{Bloch2008}%
  \BibitemOpen
  \bibfield  {author} {\bibinfo {author} {\bibfnamefont {I.}~\bibnamefont
  {Bloch}}, \bibinfo {author} {\bibfnamefont {J.}~\bibnamefont {Dalibard}}, \
  and\ \bibinfo {author} {\bibfnamefont {W.}~\bibnamefont {Zwerger}},\ }\href
  {\doibase 10.1103/RevModPhys.80.885} {\bibfield  {journal} {\bibinfo
  {journal} {Rev. Mod. Phys.}\ }\textbf {\bibinfo {volume} {80}},\ \bibinfo
  {pages} {885} (\bibinfo {year} {2008})}\BibitemShut {NoStop}%
\bibitem [{\citenamefont {Rachel}(2018)}]{Rachel2018}%
  \BibitemOpen
  \bibfield  {author} {\bibinfo {author} {\bibfnamefont {S.}~\bibnamefont
  {Rachel}},\ }\href {\doibase 10.1088/1361-6633/aad6a6} {\bibfield  {journal}
  {\bibinfo  {journal} {Reports on Progress in Physics}\ }\textbf {\bibinfo
  {volume} {81}},\ \bibinfo {pages} {116501} (\bibinfo {year}
  {2018})}\BibitemShut {NoStop}%
\bibitem [{\citenamefont {Varney}\ \emph {et~al.}(2010)\citenamefont {Varney},
  \citenamefont {Sun}, \citenamefont {Rigol},\ and\ \citenamefont
  {Galitski}}]{Varney2010}%
  \BibitemOpen
  \bibfield  {author} {\bibinfo {author} {\bibfnamefont {C.~N.}\ \bibnamefont
  {Varney}}, \bibinfo {author} {\bibfnamefont {K.}~\bibnamefont {Sun}},
  \bibinfo {author} {\bibfnamefont {M.}~\bibnamefont {Rigol}}, \ and\ \bibinfo
  {author} {\bibfnamefont {V.}~\bibnamefont {Galitski}},\ }\href
  {https://link.aps.org/doi/10.1103/PhysRevB.82.115125} {\bibfield  {journal}
  {\bibinfo  {journal} {Phys. Rev. B}\ }\textbf {\bibinfo {volume} {82}},\
  \bibinfo {pages} {115125} (\bibinfo {year} {2010})}\BibitemShut {NoStop}%
\bibitem [{\citenamefont {He}\ \emph {et~al.}(2011)\citenamefont {He},
  \citenamefont {Zong}, \citenamefont {Kou}, \citenamefont {Liang},\ and\
  \citenamefont {Feng}}]{He2011}%
  \BibitemOpen
  \bibfield  {author} {\bibinfo {author} {\bibfnamefont {J.}~\bibnamefont
  {He}}, \bibinfo {author} {\bibfnamefont {Y.-H.}\ \bibnamefont {Zong}},
  \bibinfo {author} {\bibfnamefont {S.-P.}\ \bibnamefont {Kou}}, \bibinfo
  {author} {\bibfnamefont {Y.}~\bibnamefont {Liang}}, \ and\ \bibinfo {author}
  {\bibfnamefont {S.}~\bibnamefont {Feng}},\ }\href
  {https://link.aps.org/doi/10.1103/PhysRevB.84.035127} {\bibfield  {journal}
  {\bibinfo  {journal} {Phys. Rev. B}\ }\textbf {\bibinfo {volume} {84}},\
  \bibinfo {pages} {035127} (\bibinfo {year} {2011})}\BibitemShut {NoStop}%
\bibitem [{\citenamefont {Vanhala}\ \emph {et~al.}(2016)\citenamefont
  {Vanhala}, \citenamefont {Siro}, \citenamefont {Liang}, \citenamefont
  {Troyer}, \citenamefont {Harju},\ and\ \citenamefont
  {Törmä}}]{Vanhala2016}%
  \BibitemOpen
  \bibfield  {author} {\bibinfo {author} {\bibfnamefont {T.~I.}\ \bibnamefont
  {Vanhala}}, \bibinfo {author} {\bibfnamefont {T.}~\bibnamefont {Siro}},
  \bibinfo {author} {\bibfnamefont {L.}~\bibnamefont {Liang}}, \bibinfo
  {author} {\bibfnamefont {M.}~\bibnamefont {Troyer}}, \bibinfo {author}
  {\bibfnamefont {A.}~\bibnamefont {Harju}}, \ and\ \bibinfo {author}
  {\bibfnamefont {P.}~\bibnamefont {Törmä}},\ }\href {\doibase
  10.1103/PhysRevLett.116.225305} {\bibfield  {journal} {\bibinfo  {journal}
  {Phys. Rev. Lett.}\ }\textbf {\bibinfo {volume} {116}},\ \bibinfo {pages}
  {225305} (\bibinfo {year} {2016})}\BibitemShut {NoStop}%
\bibitem [{\citenamefont {Tupitsyn}\ and\ \citenamefont
  {Prokof'ev}(2019)}]{Tupitsyn2019}%
  \BibitemOpen
  \bibfield  {author} {\bibinfo {author} {\bibfnamefont {I.~S.}\ \bibnamefont
  {Tupitsyn}}\ and\ \bibinfo {author} {\bibfnamefont {N.~V.}\ \bibnamefont
  {Prokof'ev}},\ }\href {\doibase 10.1103/PhysRevB.99.121113} {\bibfield
  {journal} {\bibinfo  {journal} {Phys. Rev. B}\ }\textbf {\bibinfo {volume}
  {99}},\ \bibinfo {pages} {121113} (\bibinfo {year} {2019})}\BibitemShut
  {NoStop}%
\bibitem [{\citenamefont {Mertz}\ \emph {et~al.}(2019)\citenamefont {Mertz},
  \citenamefont {Zantout},\ and\ \citenamefont {Valent\'{\i}}}]{Mertz2019}%
  \BibitemOpen
  \bibfield  {author} {\bibinfo {author} {\bibfnamefont {T.}~\bibnamefont
  {Mertz}}, \bibinfo {author} {\bibfnamefont {K.}~\bibnamefont {Zantout}}, \
  and\ \bibinfo {author} {\bibfnamefont {R.}~\bibnamefont {Valent\'{\i}}},\
  }\href {\doibase 10.1103/PhysRevB.100.125111} {\bibfield  {journal} {\bibinfo
   {journal} {Phys. Rev. B}\ }\textbf {\bibinfo {volume} {100}},\ \bibinfo
  {pages} {125111} (\bibinfo {year} {2019})}\BibitemShut {NoStop}%
\bibitem [{\citenamefont {Cocks}\ \emph {et~al.}(2012)\citenamefont {Cocks},
  \citenamefont {Orth}, \citenamefont {Rachel}, \citenamefont {Buchhold},
  \citenamefont {{Le Hur}},\ and\ \citenamefont {Hofstetter}}]{Cocks2012}%
  \BibitemOpen
  \bibfield  {author} {\bibinfo {author} {\bibfnamefont {D.}~\bibnamefont
  {Cocks}}, \bibinfo {author} {\bibfnamefont {P.~P.}\ \bibnamefont {Orth}},
  \bibinfo {author} {\bibfnamefont {S.}~\bibnamefont {Rachel}}, \bibinfo
  {author} {\bibfnamefont {M.}~\bibnamefont {Buchhold}}, \bibinfo {author}
  {\bibfnamefont {K.}~\bibnamefont {{Le Hur}}}, \ and\ \bibinfo {author}
  {\bibfnamefont {W.}~\bibnamefont {Hofstetter}},\ }\href {\doibase
  10.1103/PhysRevLett.109.205303} {\bibfield  {journal} {\bibinfo  {journal}
  {Phys. Rev. Lett.}\ }\textbf {\bibinfo {volume} {109}},\ \bibinfo {pages}
  {205303} (\bibinfo {year} {2012})}\BibitemShut {NoStop}%
\bibitem [{\citenamefont {Budich}\ \emph {et~al.}(2013)\citenamefont {Budich},
  \citenamefont {Trauzettel},\ and\ \citenamefont {Sangiovanni}}]{Budich2013}%
  \BibitemOpen
  \bibfield  {author} {\bibinfo {author} {\bibfnamefont {J.~C.}\ \bibnamefont
  {Budich}}, \bibinfo {author} {\bibfnamefont {B.}~\bibnamefont {Trauzettel}},
  \ and\ \bibinfo {author} {\bibfnamefont {G.}~\bibnamefont {Sangiovanni}},\
  }\href {\doibase 10.1103/PhysRevB.87.235104} {\bibfield  {journal} {\bibinfo
  {journal} {Phys. Rev. B}\ }\textbf {\bibinfo {volume} {87}},\ \bibinfo
  {pages} {235104} (\bibinfo {year} {2013})}\BibitemShut {NoStop}%
\bibitem [{\citenamefont {Amaricci}\ \emph {et~al.}(2015)\citenamefont
  {Amaricci}, \citenamefont {Budich}, \citenamefont {Capone}, \citenamefont
  {Trauzettel},\ and\ \citenamefont {Sangiovanni}}]{Amaricci2015}%
  \BibitemOpen
  \bibfield  {author} {\bibinfo {author} {\bibfnamefont {A.}~\bibnamefont
  {Amaricci}}, \bibinfo {author} {\bibfnamefont {J.~C.}\ \bibnamefont
  {Budich}}, \bibinfo {author} {\bibfnamefont {M.}~\bibnamefont {Capone}},
  \bibinfo {author} {\bibfnamefont {B.}~\bibnamefont {Trauzettel}}, \ and\
  \bibinfo {author} {\bibfnamefont {G.}~\bibnamefont {Sangiovanni}},\ }\href
  {\doibase 10.1103/PhysRevLett.114.185701} {\bibfield  {journal} {\bibinfo
  {journal} {Phys. Rev. Lett.}\ }\textbf {\bibinfo {volume} {114}},\ \bibinfo
  {pages} {185701} (\bibinfo {year} {2015})}\BibitemShut {NoStop}%
\bibitem [{\citenamefont {Jiang}\ \emph {et~al.}(2018)\citenamefont {Jiang},
  \citenamefont {Zhou}, \citenamefont {Dai},\ and\ \citenamefont
  {Wang}}]{Jiang2018}%
  \BibitemOpen
  \bibfield  {author} {\bibinfo {author} {\bibfnamefont {K.}~\bibnamefont
  {Jiang}}, \bibinfo {author} {\bibfnamefont {S.}~\bibnamefont {Zhou}},
  \bibinfo {author} {\bibfnamefont {X.}~\bibnamefont {Dai}}, \ and\ \bibinfo
  {author} {\bibfnamefont {Z.}~\bibnamefont {Wang}},\ }\href
  {https://link.aps.org/doi/10.1103/PhysRevLett.120.157205} {\bibfield
  {journal} {\bibinfo  {journal} {Phys. Rev. Lett.}\ }\textbf {\bibinfo
  {volume} {120}},\ \bibinfo {pages} {157205} (\bibinfo {year}
  {2018})}\BibitemShut {NoStop}%
\bibitem [{\citenamefont {{Amaricci A.}}\ \emph {et~al.}(2016)\citenamefont
  {{Amaricci A.}}, \citenamefont {{Budich J. C.}}, \citenamefont {{Capone M.}},
  \citenamefont {{Trauzettel B.}},\ and\ \citenamefont {{Sangiovanni
  G.}}}]{Amaricci2016}%
  \BibitemOpen
  \bibfield  {author} {\bibinfo {author} {\bibnamefont {{Amaricci A.}}},
  \bibinfo {author} {\bibnamefont {{Budich J. C.}}}, \bibinfo {author}
  {\bibnamefont {{Capone M.}}}, \bibinfo {author} {\bibnamefont {{Trauzettel
  B.}}}, \ and\ \bibinfo {author} {\bibnamefont {{Sangiovanni G.}}},\ }\href
  {\doibase 10.1103/PhysRevB.93.235112} {\bibfield  {journal} {\bibinfo
  {journal} {Phys. Rev. B}\ }\textbf {\bibinfo {volume} {93}},\ \bibinfo
  {pages} {235112} (\bibinfo {year} {2016})}\BibitemShut {NoStop}%
\bibitem [{\citenamefont {Irsigler}\ \emph {et~al.}(2020)\citenamefont
  {Irsigler}, \citenamefont {Zheng}, \citenamefont {Grusdt},\ and\
  \citenamefont {Hofstetter}}]{Irsigler2020}%
  \BibitemOpen
  \bibfield  {author} {\bibinfo {author} {\bibfnamefont {B.}~\bibnamefont
  {Irsigler}}, \bibinfo {author} {\bibfnamefont {J.-H.}\ \bibnamefont {Zheng}},
  \bibinfo {author} {\bibfnamefont {F.}~\bibnamefont {Grusdt}}, \ and\ \bibinfo
  {author} {\bibfnamefont {W.}~\bibnamefont {Hofstetter}},\ }\href {\doibase
  10.1103/PhysRevResearch.2.013299} {\bibfield  {journal} {\bibinfo  {journal}
  {Phys. Rev. Research}\ }\textbf {\bibinfo {volume} {2}},\ \bibinfo {pages}
  {013299} (\bibinfo {year} {2020})}\BibitemShut {NoStop}%
\bibitem [{\citenamefont {Hafez-Torbati}\ \emph {et~al.}(2020)\citenamefont
  {Hafez-Torbati}, \citenamefont {Zheng}, \citenamefont {Irsigler},\ and\
  \citenamefont {Hofstetter}}]{Hafez-Torbati2020}%
  \BibitemOpen
  \bibfield  {author} {\bibinfo {author} {\bibfnamefont {M.}~\bibnamefont
  {Hafez-Torbati}}, \bibinfo {author} {\bibfnamefont {J.-H.}\ \bibnamefont
  {Zheng}}, \bibinfo {author} {\bibfnamefont {B.}~\bibnamefont {Irsigler}}, \
  and\ \bibinfo {author} {\bibfnamefont {W.}~\bibnamefont {Hofstetter}},\
  }\href {\doibase 10.1103/PhysRevB.101.245159} {\bibfield  {journal} {\bibinfo
   {journal} {Phys. Rev. B}\ }\textbf {\bibinfo {volume} {101}},\ \bibinfo
  {pages} {245159} (\bibinfo {year} {2020})}\BibitemShut {NoStop}%
\bibitem [{\citenamefont {Balents}(2010)}]{Balents2010}%
  \BibitemOpen
  \bibfield  {author} {\bibinfo {author} {\bibfnamefont {L.}~\bibnamefont
  {Balents}},\ }\href {https://doi.org/10.1038/nature08917} {\bibfield
  {journal} {\bibinfo  {journal} {Nature}\ }\textbf {\bibinfo {volume} {464}},\
  \bibinfo {pages} {199–208} (\bibinfo {year} {2010})}\BibitemShut {NoStop}%
\bibitem [{\citenamefont {Radi\ifmmode~\acute{c}\else \'{c}\fi{}}\ \emph
  {et~al.}(2012)\citenamefont {Radi\ifmmode~\acute{c}\else \'{c}\fi{}},
  \citenamefont {Di~Ciolo}, \citenamefont {Sun},\ and\ \citenamefont
  {Galitski}}]{Radic2012}%
  \BibitemOpen
  \bibfield  {author} {\bibinfo {author} {\bibfnamefont {J.}~\bibnamefont
  {Radi\ifmmode~\acute{c}\else \'{c}\fi{}}}, \bibinfo {author} {\bibfnamefont
  {A.}~\bibnamefont {Di~Ciolo}}, \bibinfo {author} {\bibfnamefont
  {K.}~\bibnamefont {Sun}}, \ and\ \bibinfo {author} {\bibfnamefont
  {V.}~\bibnamefont {Galitski}},\ }\href {\doibase
  10.1103/PhysRevLett.109.085303} {\bibfield  {journal} {\bibinfo  {journal}
  {Phys. Rev. Lett.}\ }\textbf {\bibinfo {volume} {109}},\ \bibinfo {pages}
  {085303} (\bibinfo {year} {2012})}\BibitemShut {NoStop}%
\bibitem [{\citenamefont {Arun}\ \emph {et~al.}(2016)\citenamefont {Arun},
  \citenamefont {Sohal}, \citenamefont {Hickey},\ and\ \citenamefont
  {Paramekanti}}]{Arun2016}%
  \BibitemOpen
  \bibfield  {author} {\bibinfo {author} {\bibfnamefont {V.~S.}\ \bibnamefont
  {Arun}}, \bibinfo {author} {\bibfnamefont {R.}~\bibnamefont {Sohal}},
  \bibinfo {author} {\bibfnamefont {C.}~\bibnamefont {Hickey}}, \ and\ \bibinfo
  {author} {\bibfnamefont {A.}~\bibnamefont {Paramekanti}},\ }\href {\doibase
  10.1103/PhysRevB.93.115110} {\bibfield  {journal} {\bibinfo  {journal} {Phys.
  Rev. B}\ }\textbf {\bibinfo {volume} {93}},\ \bibinfo {pages} {115110}
  (\bibinfo {year} {2016})}\BibitemShut {NoStop}%
\bibitem [{\citenamefont {Irsigler}\ \emph
  {et~al.}(2019{\natexlab{a}})\citenamefont {Irsigler}, \citenamefont {Zheng},
  \citenamefont {Hafez-Torbati},\ and\ \citenamefont
  {Hofstetter}}]{Irsigler2019a}%
  \BibitemOpen
  \bibfield  {author} {\bibinfo {author} {\bibfnamefont {B.}~\bibnamefont
  {Irsigler}}, \bibinfo {author} {\bibfnamefont {J.-H.}\ \bibnamefont {Zheng}},
  \bibinfo {author} {\bibfnamefont {M.}~\bibnamefont {Hafez-Torbati}}, \ and\
  \bibinfo {author} {\bibfnamefont {W.}~\bibnamefont {Hofstetter}},\ }\href
  {\doibase 10.1103/PhysRevA.99.043628} {\bibfield  {journal} {\bibinfo
  {journal} {Phys. Rev. A}\ }\textbf {\bibinfo {volume} {99}},\ \bibinfo
  {pages} {043628} (\bibinfo {year} {2019}{\natexlab{a}})}\BibitemShut
  {NoStop}%
\bibitem [{\citenamefont {Hatsugai}\ and\ \citenamefont
  {Kohmoto}(1990)}]{Hatsugai1990}%
  \BibitemOpen
  \bibfield  {author} {\bibinfo {author} {\bibfnamefont {Y.}~\bibnamefont
  {Hatsugai}}\ and\ \bibinfo {author} {\bibfnamefont {M.}~\bibnamefont
  {Kohmoto}},\ }\href {\doibase 10.1103/PhysRevB.42.8282} {\bibfield  {journal}
  {\bibinfo  {journal} {Phys. Rev. B}\ }\textbf {\bibinfo {volume} {42}},\
  \bibinfo {pages} {8282} (\bibinfo {year} {1990})}\BibitemShut {NoStop}%
\bibitem [{\citenamefont {Han}\ \emph {et~al.}(1994)\citenamefont {Han},
  \citenamefont {Thouless}, \citenamefont {Hiramoto},\ and\ \citenamefont
  {Kohmoto}}]{Han1994}%
  \BibitemOpen
  \bibfield  {author} {\bibinfo {author} {\bibfnamefont {J.~H.}\ \bibnamefont
  {Han}}, \bibinfo {author} {\bibfnamefont {D.~J.}\ \bibnamefont {Thouless}},
  \bibinfo {author} {\bibfnamefont {H.}~\bibnamefont {Hiramoto}}, \ and\
  \bibinfo {author} {\bibfnamefont {M.}~\bibnamefont {Kohmoto}},\ }\href
  {\doibase 10.1103/PhysRevB.50.11365} {\bibfield  {journal} {\bibinfo
  {journal} {Phys. Rev. B}\ }\textbf {\bibinfo {volume} {50}},\ \bibinfo
  {pages} {11365} (\bibinfo {year} {1994})}\BibitemShut {NoStop}%
\bibitem [{\citenamefont {Thouless}(1983)}]{Thouless1983}%
  \BibitemOpen
  \bibfield  {author} {\bibinfo {author} {\bibfnamefont {D.~J.}\ \bibnamefont
  {Thouless}},\ }\href {\doibase 10.1103/PhysRevB.28.4272} {\bibfield
  {journal} {\bibinfo  {journal} {Phys. Rev. B}\ }\textbf {\bibinfo {volume}
  {28}},\ \bibinfo {pages} {4272} (\bibinfo {year} {1983})}\BibitemShut
  {NoStop}%
\bibitem [{\citenamefont {Beugeling}\ \emph {et~al.}(2012)\citenamefont
  {Beugeling}, \citenamefont {Everts},\ and\ \citenamefont
  {Morais~Smith}}]{Beugeling2012}%
  \BibitemOpen
  \bibfield  {author} {\bibinfo {author} {\bibfnamefont {W.}~\bibnamefont
  {Beugeling}}, \bibinfo {author} {\bibfnamefont {J.~C.}\ \bibnamefont
  {Everts}}, \ and\ \bibinfo {author} {\bibfnamefont {C.}~\bibnamefont
  {Morais~Smith}},\ }\href {\doibase 10.1103/PhysRevB.86.195129} {\bibfield
  {journal} {\bibinfo  {journal} {Phys. Rev. B}\ }\textbf {\bibinfo {volume}
  {86}},\ \bibinfo {pages} {195129} (\bibinfo {year} {2012})}\BibitemShut
  {NoStop}%
\bibitem [{\citenamefont {Di~Liberto}\ \emph {et~al.}(2011)\citenamefont
  {Di~Liberto}, \citenamefont {Tieleman}, \citenamefont {Branchina},\ and\
  \citenamefont {Smith}}]{DiLiberto2011}%
  \BibitemOpen
  \bibfield  {author} {\bibinfo {author} {\bibfnamefont {M.}~\bibnamefont
  {Di~Liberto}}, \bibinfo {author} {\bibfnamefont {O.}~\bibnamefont
  {Tieleman}}, \bibinfo {author} {\bibfnamefont {V.}~\bibnamefont {Branchina}},
  \ and\ \bibinfo {author} {\bibfnamefont {C.~M.}\ \bibnamefont {Smith}},\
  }\href {\doibase 10.1103/PhysRevA.84.013607} {\bibfield  {journal} {\bibinfo
  {journal} {Phys. Rev. A}\ }\textbf {\bibinfo {volume} {84}},\ \bibinfo
  {pages} {013607} (\bibinfo {year} {2011})}\BibitemShut {NoStop}%
\bibitem [{\citenamefont {Gerbier}\ and\ \citenamefont
  {Dalibard}(2010)}]{Gerbier2010}%
  \BibitemOpen
  \bibfield  {author} {\bibinfo {author} {\bibfnamefont {F.}~\bibnamefont
  {Gerbier}}\ and\ \bibinfo {author} {\bibfnamefont {J.}~\bibnamefont
  {Dalibard}},\ }\href {\doibase 10.1088/1367-2630/12/3/033007} {\bibfield
  {journal} {\bibinfo  {journal} {New Journal of Physics}\ }\textbf {\bibinfo
  {volume} {12}},\ \bibinfo {pages} {033007} (\bibinfo {year}
  {2010})}\BibitemShut {NoStop}%
\bibitem [{\citenamefont {Goldman}\ \emph {et~al.}(2010)\citenamefont
  {Goldman}, \citenamefont {Satija}, \citenamefont {Nikolic}, \citenamefont
  {Bermudez}, \citenamefont {Martin-Delgado}, \citenamefont {Lewenstein},\ and\
  \citenamefont {Spielman}}]{Goldman2010}%
  \BibitemOpen
  \bibfield  {author} {\bibinfo {author} {\bibfnamefont {N.}~\bibnamefont
  {Goldman}}, \bibinfo {author} {\bibfnamefont {I.}~\bibnamefont {Satija}},
  \bibinfo {author} {\bibfnamefont {P.}~\bibnamefont {Nikolic}}, \bibinfo
  {author} {\bibfnamefont {A.}~\bibnamefont {Bermudez}}, \bibinfo {author}
  {\bibfnamefont {M.~A.}\ \bibnamefont {Martin-Delgado}}, \bibinfo {author}
  {\bibfnamefont {M.}~\bibnamefont {Lewenstein}}, \ and\ \bibinfo {author}
  {\bibfnamefont {I.~B.}\ \bibnamefont {Spielman}},\ }\href {\doibase
  10.1103/PhysRevLett.105.255302} {\bibfield  {journal} {\bibinfo  {journal}
  {Phys. Rev. Lett.}\ }\textbf {\bibinfo {volume} {105}},\ \bibinfo {pages}
  {255302} (\bibinfo {year} {2010})}\BibitemShut {NoStop}%
\bibitem [{\citenamefont {Zheng}\ \emph {et~al.}(2019)\citenamefont {Zheng},
  \citenamefont {Qin},\ and\ \citenamefont {Hofstetter}}]{Zheng2019}%
  \BibitemOpen
  \bibfield  {author} {\bibinfo {author} {\bibfnamefont {J.-H.}\ \bibnamefont
  {Zheng}}, \bibinfo {author} {\bibfnamefont {T.}~\bibnamefont {Qin}}, \ and\
  \bibinfo {author} {\bibfnamefont {W.}~\bibnamefont {Hofstetter}},\ }\href
  {\doibase 10.1103/PhysRevB.99.125138} {\bibfield  {journal} {\bibinfo
  {journal} {Phys. Rev. B}\ }\textbf {\bibinfo {volume} {99}},\ \bibinfo
  {pages} {125138} (\bibinfo {year} {2019})}\BibitemShut {NoStop}%
\bibitem [{\citenamefont {Hafez-Torbati}\ and\ \citenamefont
  {Uhrig}(2016)}]{Hafez-Torbati2016}%
  \BibitemOpen
  \bibfield  {author} {\bibinfo {author} {\bibfnamefont {M.}~\bibnamefont
  {Hafez-Torbati}}\ and\ \bibinfo {author} {\bibfnamefont {G.~S.}\ \bibnamefont
  {Uhrig}},\ }\href {\doibase 10.1103/PhysRevB.93.195128} {\bibfield  {journal}
  {\bibinfo  {journal} {Phys. Rev. B}\ }\textbf {\bibinfo {volume} {93}},\
  \bibinfo {pages} {195128} (\bibinfo {year} {2016})}\BibitemShut {NoStop}%
\bibitem [{\citenamefont {Kancharla}\ and\ \citenamefont
  {Dagotto}(2007)}]{Kancharla2007}%
  \BibitemOpen
  \bibfield  {author} {\bibinfo {author} {\bibfnamefont {S.~S.}\ \bibnamefont
  {Kancharla}}\ and\ \bibinfo {author} {\bibfnamefont {E.}~\bibnamefont
  {Dagotto}},\ }\href {\doibase 10.1103/PhysRevLett.98.016402} {\bibfield
  {journal} {\bibinfo  {journal} {Phys. Rev. Lett.}\ }\textbf {\bibinfo
  {volume} {98}},\ \bibinfo {pages} {016402} (\bibinfo {year}
  {2007})}\BibitemShut {NoStop}%
\bibitem [{\citenamefont {Paris}\ \emph {et~al.}(2007)\citenamefont {Paris},
  \citenamefont {Bouadim}, \citenamefont {Hébert}, \citenamefont {Batrouni},\
  and\ \citenamefont {Scalettar}}]{Paris2007}%
  \BibitemOpen
  \bibfield  {author} {\bibinfo {author} {\bibfnamefont {N.}~\bibnamefont
  {Paris}}, \bibinfo {author} {\bibfnamefont {K.}~\bibnamefont {Bouadim}},
  \bibinfo {author} {\bibfnamefont {F.}~\bibnamefont {Hébert}}, \bibinfo
  {author} {\bibfnamefont {G.~G.}\ \bibnamefont {Batrouni}}, \ and\ \bibinfo
  {author} {\bibfnamefont {R.~T.}\ \bibnamefont {Scalettar}},\ }\href {\doibase
  10.1103/PhysRevLett.98.046403} {\bibfield  {journal} {\bibinfo  {journal}
  {Phys. Rev. Lett.}\ }\textbf {\bibinfo {volume} {98}},\ \bibinfo {pages}
  {046403} (\bibinfo {year} {2007})}\BibitemShut {NoStop}%
\bibitem [{\citenamefont {Wang}\ \emph {et~al.}(2020)\citenamefont {Wang},
  \citenamefont {Zhang}, \citenamefont {Ma}, \citenamefont {Chen},
  \citenamefont {Liang},\ and\ \citenamefont {Ma}}]{Wang2020}%
  \BibitemOpen
  \bibfield  {author} {\bibinfo {author} {\bibfnamefont {J.}~\bibnamefont
  {Wang}}, \bibinfo {author} {\bibfnamefont {L.}~\bibnamefont {Zhang}},
  \bibinfo {author} {\bibfnamefont {R.}~\bibnamefont {Ma}}, \bibinfo {author}
  {\bibfnamefont {Q.}~\bibnamefont {Chen}}, \bibinfo {author} {\bibfnamefont
  {Y.}~\bibnamefont {Liang}}, \ and\ \bibinfo {author} {\bibfnamefont
  {T.}~\bibnamefont {Ma}},\ }\href {\doibase 10.1103/PhysRevB.101.245161}
  {\bibfield  {journal} {\bibinfo  {journal} {Phys. Rev. B}\ }\textbf {\bibinfo
  {volume} {101}},\ \bibinfo {pages} {245161} (\bibinfo {year}
  {2020})}\BibitemShut {NoStop}%
\bibitem [{\citenamefont {Lin}\ \emph {et~al.}(2015)\citenamefont {Lin},
  \citenamefont {Liu}, \citenamefont {Tao},\ and\ \citenamefont
  {Liu}}]{Lin2015}%
  \BibitemOpen
  \bibfield  {author} {\bibinfo {author} {\bibfnamefont {H.-F.}\ \bibnamefont
  {Lin}}, \bibinfo {author} {\bibfnamefont {H.-D.}\ \bibnamefont {Liu}},
  \bibinfo {author} {\bibfnamefont {H.-S.}\ \bibnamefont {Tao}}, \ and\
  \bibinfo {author} {\bibfnamefont {W.-M.}\ \bibnamefont {Liu}},\ }\href
  {https://doi.org/10.1038/srep09810} {\bibfield  {journal} {\bibinfo
  {journal} {Scientific Reports}\ }\textbf {\bibinfo {volume} {5}},\ \bibinfo
  {pages} {9810} (\bibinfo {year} {2015})}\BibitemShut {NoStop}%
\bibitem [{\citenamefont {Ebrahimkhas}(2011)}]{Ebrahimkhas2011}%
  \BibitemOpen
  \bibfield  {author} {\bibinfo {author} {\bibfnamefont {M.}~\bibnamefont
  {Ebrahimkhas}},\ }\href {\doibase 10.1016/j.physleta.2011.07.019} {\bibfield
  {journal} {\bibinfo  {journal} {Physics Letters A}\ }\textbf {\bibinfo
  {volume} {375}},\ \bibinfo {pages} {3223} (\bibinfo {year}
  {2011})}\BibitemShut {NoStop}%
\bibitem [{\citenamefont {Shahbazy}\ and\ \citenamefont
  {Ebrahimkhas}(2019)}]{Shahbazy2019}%
  \BibitemOpen
  \bibfield  {author} {\bibinfo {author} {\bibfnamefont {A.}~\bibnamefont
  {Shahbazy}}\ and\ \bibinfo {author} {\bibfnamefont {M.}~\bibnamefont
  {Ebrahimkhas}},\ }\href {\doibase 10.1016/j.cjph.2019.01.013} {\bibfield
  {journal} {\bibinfo  {journal} {Chinese Journal of Physics}\ }\textbf
  {\bibinfo {volume} {58}},\ \bibinfo {pages} {273} (\bibinfo {year}
  {2019})}\BibitemShut {NoStop}%
\bibitem [{\citenamefont {Kotliar}\ and\ \citenamefont
  {Vollhardt}(2004)}]{Kotliar2004}%
  \BibitemOpen
  \bibfield  {author} {\bibinfo {author} {\bibfnamefont {G.}~\bibnamefont
  {Kotliar}}\ and\ \bibinfo {author} {\bibfnamefont {D.}~\bibnamefont
  {Vollhardt}},\ }\href {\doibase 10.1063/1.1712502} {\bibfield  {journal}
  {\bibinfo  {journal} {Physics Today}\ }\textbf {\bibinfo {volume} {57}},\
  \bibinfo {pages} {53} (\bibinfo {year} {2004})}\BibitemShut {NoStop}%
\bibitem [{\citenamefont {Georges}\ \emph {et~al.}(1996)\citenamefont
  {Georges}, \citenamefont {Kotliar}, \citenamefont {Krauth},\ and\
  \citenamefont {Rozenberg}}]{Georges1996}%
  \BibitemOpen
  \bibfield  {author} {\bibinfo {author} {\bibfnamefont {A.}~\bibnamefont
  {Georges}}, \bibinfo {author} {\bibfnamefont {G.}~\bibnamefont {Kotliar}},
  \bibinfo {author} {\bibfnamefont {W.}~\bibnamefont {Krauth}}, \ and\ \bibinfo
  {author} {\bibfnamefont {M.~J.}\ \bibnamefont {Rozenberg}},\ }\href {\doibase
  10.1103/RevModPhys.68.13} {\bibfield  {journal} {\bibinfo  {journal} {Rev.
  Mod. Phys.}\ }\textbf {\bibinfo {volume} {68}},\ \bibinfo {pages} {13}
  (\bibinfo {year} {1996})}\BibitemShut {NoStop}%
\bibitem [{\citenamefont {Metzner}\ and\ \citenamefont
  {Vollhardt}(1989)}]{Metzner1989}%
  \BibitemOpen
  \bibfield  {author} {\bibinfo {author} {\bibfnamefont {W.}~\bibnamefont
  {Metzner}}\ and\ \bibinfo {author} {\bibfnamefont {D.}~\bibnamefont
  {Vollhardt}},\ }\href {\doibase 10.1103/PhysRevLett.62.324} {\bibfield
  {journal} {\bibinfo  {journal} {Phys. Rev. Lett.}\ }\textbf {\bibinfo
  {volume} {62}},\ \bibinfo {pages} {324} (\bibinfo {year} {1989})}\BibitemShut
  {NoStop}%
\bibitem [{\citenamefont {Potthoff}\ and\ \citenamefont
  {Nolting}(1999)}]{Potthoff1999}%
  \BibitemOpen
  \bibfield  {author} {\bibinfo {author} {\bibfnamefont {M.}~\bibnamefont
  {Potthoff}}\ and\ \bibinfo {author} {\bibfnamefont {W.}~\bibnamefont
  {Nolting}},\ }\href {\doibase 10.1103/PhysRevB.59.2549} {\bibfield  {journal}
  {\bibinfo  {journal} {Phys. Rev. B}\ }\textbf {\bibinfo {volume} {59}},\
  \bibinfo {pages} {2549} (\bibinfo {year} {1999})}\BibitemShut {NoStop}%
\bibitem [{\citenamefont {Song}\ \emph {et~al.}(2008)\citenamefont {Song},
  \citenamefont {Wortis},\ and\ \citenamefont {Atkinson}}]{Song2008}%
  \BibitemOpen
  \bibfield  {author} {\bibinfo {author} {\bibfnamefont {Y.}~\bibnamefont
  {Song}}, \bibinfo {author} {\bibfnamefont {R.}~\bibnamefont {Wortis}}, \ and\
  \bibinfo {author} {\bibfnamefont {W.~A.}\ \bibnamefont {Atkinson}},\ }\href
  {\doibase 10.1103/PhysRevB.77.054202} {\bibfield  {journal} {\bibinfo
  {journal} {Phys. Rev. B}\ }\textbf {\bibinfo {volume} {77}},\ \bibinfo
  {pages} {054202} (\bibinfo {year} {2008})}\BibitemShut {NoStop}%
\bibitem [{\citenamefont {Snoek}\ \emph {et~al.}(2008)\citenamefont {Snoek},
  \citenamefont {Titvinidze}, \citenamefont {T{\H{o}}ke}, \citenamefont
  {Byczuk},\ and\ \citenamefont {Hofstetter}}]{Snoek2008}%
  \BibitemOpen
  \bibfield  {author} {\bibinfo {author} {\bibfnamefont {M.}~\bibnamefont
  {Snoek}}, \bibinfo {author} {\bibfnamefont {I.}~\bibnamefont {Titvinidze}},
  \bibinfo {author} {\bibfnamefont {C.}~\bibnamefont {T{\H{o}}ke}}, \bibinfo
  {author} {\bibfnamefont {K.}~\bibnamefont {Byczuk}}, \ and\ \bibinfo {author}
  {\bibfnamefont {W.}~\bibnamefont {Hofstetter}},\ }\href {\doibase
  10.1088/1367-2630/10/9/093008} {\bibfield  {journal} {\bibinfo  {journal}
  {New Journal of Physics}\ }\textbf {\bibinfo {volume} {10}},\ \bibinfo
  {pages} {093008} (\bibinfo {year} {2008})}\BibitemShut {NoStop}%
\bibitem [{\citenamefont {Orth}\ \emph {et~al.}(2013)\citenamefont {Orth},
  \citenamefont {Cocks}, \citenamefont {Rachel}, \citenamefont {Buchhold},
  \citenamefont {Hur},\ and\ \citenamefont {Hofstetter}}]{Orth2013}%
  \BibitemOpen
  \bibfield  {author} {\bibinfo {author} {\bibfnamefont {P.~P.}\ \bibnamefont
  {Orth}}, \bibinfo {author} {\bibfnamefont {D.}~\bibnamefont {Cocks}},
  \bibinfo {author} {\bibfnamefont {S.}~\bibnamefont {Rachel}}, \bibinfo
  {author} {\bibfnamefont {M.}~\bibnamefont {Buchhold}}, \bibinfo {author}
  {\bibfnamefont {K.~L.}\ \bibnamefont {Hur}}, \ and\ \bibinfo {author}
  {\bibfnamefont {W.}~\bibnamefont {Hofstetter}},\ }\href
  {http://stacks.iop.org/0953-4075/46/i=13/a=134004} {\bibfield  {journal}
  {\bibinfo  {journal} {Journal of Physics B: Atomic, Molecular and Optical
  Physics}\ }\textbf {\bibinfo {volume} {46}},\ \bibinfo {pages} {134004}
  (\bibinfo {year} {2013})}\BibitemShut {NoStop}%
\bibitem [{\citenamefont {Hafez-Torbati}\ and\ \citenamefont
  {Hofstetter}(2019)}]{Hafez-Torbati2019}%
  \BibitemOpen
  \bibfield  {author} {\bibinfo {author} {\bibfnamefont {M.}~\bibnamefont
  {Hafez-Torbati}}\ and\ \bibinfo {author} {\bibfnamefont {W.}~\bibnamefont
  {Hofstetter}},\ }\href {https://link.aps.org/doi/10.1103/PhysRevB.100.035133}
  {\bibfield  {journal} {\bibinfo  {journal} {Phys. Rev. B}\ }\textbf {\bibinfo
  {volume} {100}},\ \bibinfo {pages} {035133} (\bibinfo {year}
  {2019})}\BibitemShut {NoStop}%
\bibitem [{\citenamefont {{Valli Angelo}}\ \emph {et~al.}(2018)\citenamefont
  {{Valli Angelo}}, \citenamefont {{Amaricci Adriano}}, \citenamefont {{Brosco
  Valentina}},\ and\ \citenamefont {{Capone Massimo}}}]{Valli2018}%
  \BibitemOpen
  \bibfield  {author} {\bibinfo {author} {\bibnamefont {{Valli Angelo}}},
  \bibinfo {author} {\bibnamefont {{Amaricci Adriano}}}, \bibinfo {author}
  {\bibnamefont {{Brosco Valentina}}}, \ and\ \bibinfo {author} {\bibnamefont
  {{Capone Massimo}}},\ }\href {\doibase 10.1021/acs.nanolett.8b00453}
  {\bibfield  {journal} {\bibinfo  {journal} {Nano Letters}\ }\textbf {\bibinfo
  {volume} {18}},\ \bibinfo {pages} {2158–2164} (\bibinfo {year}
  {2018})}\BibitemShut {NoStop}%
\bibitem [{\citenamefont {Irsigler}\ \emph
  {et~al.}(2019{\natexlab{b}})\citenamefont {Irsigler}, \citenamefont {Zheng},\
  and\ \citenamefont {Hofstetter}}]{Irsigler2019}%
  \BibitemOpen
  \bibfield  {author} {\bibinfo {author} {\bibfnamefont {B.}~\bibnamefont
  {Irsigler}}, \bibinfo {author} {\bibfnamefont {J.-H.}\ \bibnamefont {Zheng}},
  \ and\ \bibinfo {author} {\bibfnamefont {W.}~\bibnamefont {Hofstetter}},\
  }\href {\doibase 10.1103/PhysRevLett.122.010406} {\bibfield  {journal}
  {\bibinfo  {journal} {Phys. Rev. Lett.}\ }\textbf {\bibinfo {volume} {122}},\
  \bibinfo {pages} {010406} (\bibinfo {year} {2019}{\natexlab{b}})}\BibitemShut
  {NoStop}%
\bibitem [{\citenamefont {Amaricci}\ \emph {et~al.}(2018)\citenamefont
  {Amaricci}, \citenamefont {Valli}, \citenamefont {Sangiovanni}, \citenamefont
  {Trauzettel},\ and\ \citenamefont {Capone}}]{Amaricci2018}%
  \BibitemOpen
  \bibfield  {author} {\bibinfo {author} {\bibfnamefont {A.}~\bibnamefont
  {Amaricci}}, \bibinfo {author} {\bibfnamefont {A.}~\bibnamefont {Valli}},
  \bibinfo {author} {\bibfnamefont {G.}~\bibnamefont {Sangiovanni}}, \bibinfo
  {author} {\bibfnamefont {B.}~\bibnamefont {Trauzettel}}, \ and\ \bibinfo
  {author} {\bibfnamefont {M.}~\bibnamefont {Capone}},\ }\href {\doibase
  10.1103/PhysRevB.98.045133} {\bibfield  {journal} {\bibinfo  {journal} {Phys.
  Rev. B}\ }\textbf {\bibinfo {volume} {98}},\ \bibinfo {pages} {045133}
  (\bibinfo {year} {2018})}\BibitemShut {NoStop}%
\bibitem [{\citenamefont {Hafez-Torbati}\ and\ \citenamefont
  {Hofstetter}(2018)}]{Hafez-Torbati2018}%
  \BibitemOpen
  \bibfield  {author} {\bibinfo {author} {\bibfnamefont {M.}~\bibnamefont
  {Hafez-Torbati}}\ and\ \bibinfo {author} {\bibfnamefont {W.}~\bibnamefont
  {Hofstetter}},\ }\href {\doibase 10.1103/PhysRevB.98.245131} {\bibfield
  {journal} {\bibinfo  {journal} {Phys. Rev. B}\ }\textbf {\bibinfo {volume}
  {98}},\ \bibinfo {pages} {245131} (\bibinfo {year} {2018})}\BibitemShut
  {NoStop}%
\bibitem [{\citenamefont {Caffarel}\ and\ \citenamefont
  {Krauth}(1994)}]{Caffarel1994}%
  \BibitemOpen
  \bibfield  {author} {\bibinfo {author} {\bibfnamefont {M.}~\bibnamefont
  {Caffarel}}\ and\ \bibinfo {author} {\bibfnamefont {W.}~\bibnamefont
  {Krauth}},\ }\href {https://link.aps.org/doi/10.1103/PhysRevLett.72.1545}
  {\bibfield  {journal} {\bibinfo  {journal} {Phys. Rev. Lett.}\ }\textbf
  {\bibinfo {volume} {72}},\ \bibinfo {pages} {1545} (\bibinfo {year}
  {1994})}\BibitemShut {NoStop}%
\bibitem [{\citenamefont {Wang}\ and\ \citenamefont {Zhang}(2012)}]{Wang2012}%
  \BibitemOpen
  \bibfield  {author} {\bibinfo {author} {\bibfnamefont {Z.}~\bibnamefont
  {Wang}}\ and\ \bibinfo {author} {\bibfnamefont {S.-C.}\ \bibnamefont
  {Zhang}},\ }\href {\doibase 10.1103/PhysRevX.2.031008} {\bibfield  {journal}
  {\bibinfo  {journal} {Phys. Rev. X}\ }\textbf {\bibinfo {volume} {2}},\
  \bibinfo {pages} {031008} (\bibinfo {year} {2012})}\BibitemShut {NoStop}%
\bibitem [{zer()}]{zeroself}%
  \BibitemOpen
  \href@noop {} {\bibinfo  {journal} {We find that the real-part of the
  self-energy at the smallest (in absolute value) Matsubara frequency
  accurately describes the zero-frequency self-energy obtained by a polynomial
  fit}\ }\BibitemShut {NoStop}%
\bibitem [{\citenamefont {He}\ \emph {et~al.}(2016)\citenamefont {He},
  \citenamefont {Wu}, \citenamefont {Meng},\ and\ \citenamefont {Lu}}]{He2016}%
  \BibitemOpen
\bibfield  {journal} {  }\bibfield  {author} {\bibinfo {author} {\bibfnamefont
  {Y.-Y.}\ \bibnamefont {He}}, \bibinfo {author} {\bibfnamefont {H.-Q.}\
  \bibnamefont {Wu}}, \bibinfo {author} {\bibfnamefont {Z.~Y.}\ \bibnamefont
  {Meng}}, \ and\ \bibinfo {author} {\bibfnamefont {Z.-Y.}\ \bibnamefont
  {Lu}},\ }\href {\doibase 10.1103/PhysRevB.93.195164} {\bibfield  {journal}
  {\bibinfo  {journal} {Phys. Rev. B}\ }\textbf {\bibinfo {volume} {93}},\
  \bibinfo {pages} {195164} (\bibinfo {year} {2016})}\BibitemShut {NoStop}%
\bibitem [{\citenamefont {Jiang}\ \emph {et~al.}(2012)\citenamefont {Jiang},
  \citenamefont {Yao},\ and\ \citenamefont {Balents}}]{Jiang2012}%
  \BibitemOpen
  \bibfield  {author} {\bibinfo {author} {\bibfnamefont {H.-C.}\ \bibnamefont
  {Jiang}}, \bibinfo {author} {\bibfnamefont {H.}~\bibnamefont {Yao}}, \ and\
  \bibinfo {author} {\bibfnamefont {L.}~\bibnamefont {Balents}},\ }\href
  {\doibase 10.1103/PhysRevB.86.024424} {\bibfield  {journal} {\bibinfo
  {journal} {Phys. Rev. B}\ }\textbf {\bibinfo {volume} {86}},\ \bibinfo
  {pages} {024424} (\bibinfo {year} {2012})}\BibitemShut {NoStop}%
\bibitem [{\citenamefont {Hu}\ \emph {et~al.}(2013)\citenamefont {Hu},
  \citenamefont {Becca}, \citenamefont {Parola},\ and\ \citenamefont
  {Sorella}}]{Hu2013}%
  \BibitemOpen
  \bibfield  {author} {\bibinfo {author} {\bibfnamefont {W.-J.}\ \bibnamefont
  {Hu}}, \bibinfo {author} {\bibfnamefont {F.}~\bibnamefont {Becca}}, \bibinfo
  {author} {\bibfnamefont {A.}~\bibnamefont {Parola}}, \ and\ \bibinfo {author}
  {\bibfnamefont {S.}~\bibnamefont {Sorella}},\ }\href {\doibase
  10.1103/PhysRevB.88.060402} {\bibfield  {journal} {\bibinfo  {journal} {Phys.
  Rev. B}\ }\textbf {\bibinfo {volume} {88}},\ \bibinfo {pages} {060402}
  (\bibinfo {year} {2013})}\BibitemShut {NoStop}%
\bibitem [{\citenamefont {Poilblanc}\ \emph {et~al.}(2019)\citenamefont
  {Poilblanc}, \citenamefont {Mambrini},\ and\ \citenamefont
  {Capponi}}]{Poilblanc2019}%
  \BibitemOpen
  \bibfield  {author} {\bibinfo {author} {\bibfnamefont {D.}~\bibnamefont
  {Poilblanc}}, \bibinfo {author} {\bibfnamefont {M.}~\bibnamefont {Mambrini}},
  \ and\ \bibinfo {author} {\bibfnamefont {S.}~\bibnamefont {Capponi}},\ }\href
  {\doibase 10.21468/SciPostPhys.7.4.041} {\bibfield  {journal} {\bibinfo
  {journal} {SciPost Phys.}\ }\textbf {\bibinfo {volume} {7}},\ \bibinfo
  {pages} {41} (\bibinfo {year} {2019})}\BibitemShut {NoStop}%
\bibitem [{\citenamefont {Haghshenas}\ and\ \citenamefont
  {Sheng}(2018)}]{Haghshenas2018}%
  \BibitemOpen
  \bibfield  {author} {\bibinfo {author} {\bibfnamefont {R.}~\bibnamefont
  {Haghshenas}}\ and\ \bibinfo {author} {\bibfnamefont {D.~N.}\ \bibnamefont
  {Sheng}},\ }\href {\doibase 10.1103/PhysRevB.97.174408} {\bibfield  {journal}
  {\bibinfo  {journal} {Phys. Rev. B}\ }\textbf {\bibinfo {volume} {97}},\
  \bibinfo {pages} {174408} (\bibinfo {year} {2018})}\BibitemShut {NoStop}%
\bibitem [{\citenamefont {Wang}\ \emph {et~al.}(2016)\citenamefont {Wang},
  \citenamefont {Gu}, \citenamefont {Verstraete},\ and\ \citenamefont
  {Wen}}]{Wang2016}%
  \BibitemOpen
  \bibfield  {author} {\bibinfo {author} {\bibfnamefont {L.}~\bibnamefont
  {Wang}}, \bibinfo {author} {\bibfnamefont {Z.-C.}\ \bibnamefont {Gu}},
  \bibinfo {author} {\bibfnamefont {F.}~\bibnamefont {Verstraete}}, \ and\
  \bibinfo {author} {\bibfnamefont {X.-G.}\ \bibnamefont {Wen}},\ }\href
  {\doibase 10.1103/PhysRevB.94.075143} {\bibfield  {journal} {\bibinfo
  {journal} {Phys. Rev. B}\ }\textbf {\bibinfo {volume} {94}},\ \bibinfo
  {pages} {075143} (\bibinfo {year} {2016})}\BibitemShut {NoStop}%
\bibitem [{\citenamefont {Capriotti}\ and\ \citenamefont
  {Sorella}(2000)}]{Capriotti2000}%
  \BibitemOpen
  \bibfield  {author} {\bibinfo {author} {\bibfnamefont {L.}~\bibnamefont
  {Capriotti}}\ and\ \bibinfo {author} {\bibfnamefont {S.}~\bibnamefont
  {Sorella}},\ }\href {\doibase 10.1103/PhysRevLett.84.3173} {\bibfield
  {journal} {\bibinfo  {journal} {Phys. Rev. Lett.}\ }\textbf {\bibinfo
  {volume} {84}},\ \bibinfo {pages} {3173} (\bibinfo {year}
  {2000})}\BibitemShut {NoStop}%
\bibitem [{\citenamefont {Crippa}\ \emph {et~al.}(2020)\citenamefont {Crippa},
  \citenamefont {Amaricci}, \citenamefont {Wagner}, \citenamefont
  {Sangiovanni}, \citenamefont {Budich},\ and\ \citenamefont
  {Capone}}]{Crippa2020}%
  \BibitemOpen
  \bibfield  {author} {\bibinfo {author} {\bibfnamefont {L.}~\bibnamefont
  {Crippa}}, \bibinfo {author} {\bibfnamefont {A.}~\bibnamefont {Amaricci}},
  \bibinfo {author} {\bibfnamefont {N.}~\bibnamefont {Wagner}}, \bibinfo
  {author} {\bibfnamefont {G.}~\bibnamefont {Sangiovanni}}, \bibinfo {author}
  {\bibfnamefont {J.~C.}\ \bibnamefont {Budich}}, \ and\ \bibinfo {author}
  {\bibfnamefont {M.}~\bibnamefont {Capone}},\ }\href {\doibase
  10.1103/PhysRevResearch.2.012023} {\bibfield  {journal} {\bibinfo  {journal}
  {Physical Review Research}\ }\textbf {\bibinfo {volume} {2}},\ \bibinfo
  {pages} {012023} (\bibinfo {year} {2020})}\BibitemShut {NoStop}%
\bibitem [{nee()}]{neelc2}%
  \BibitemOpen
  \href@noop {} {\bibinfo  {journal} {A continuous evolution of the effective
  potentials in Fig. 3(c) across the magnetic transition implies an
  intermadiate $\mathcal{C}=2$ Neel AF to exist. However, this region is so
  narrow that we could not get any converged RDMFT data within it}\
  }\BibitemShut {NoStop}%
\bibitem [{str()}]{stripec2}%
  \BibitemOpen
\bibfield  {journal} {  }\href@noop {} {\bibinfo  {journal} {There is a narrow
  region between the QHI and the $\mathcal{C}=1$ stripe AFQHI where both spin
  components are in the topological state. This phase is not specified in the
  figure as it is very narrow, $\simeq 0.5t$, and may not survive beyond the
  DMFT approximation}\ }\BibitemShut {NoStop}%
\end{thebibliography}

%

\end{document}